\documentclass[11pt,twoside,a4paper,cmspaper,final,collab]{cms-tdr}
\def\svnVersion{fe9d20a-D}\def\svnDate{2026/09/28}\def\cmsCernNoTag{CERN-EP-2026-161}\def\cmsCernDate{\today}\def\cmsMessage{Submitted to Physical Review D}
\begin{document}\cmsNoteHeader{SUS-23-017}

\ifthenelse{\boolean{cms@external}}{
   \providecommand{\cmsLeft}{upper\xspace}
   \providecommand{\cmsRight}{lower\xspace}
   \newenvironment{scotchx}{\protect\centering\ruledtabular\tabular{ll}}{\endtabular\endruledtabular}
}{
      \providecommand{\cmsLeft}{left\xspace}
      \providecommand{\cmsRight}{right\xspace}
      \newenvironment{scotchx}{\protect\centering\tabular{p{0.3\textwidth} p{0.65\textwidth}}}{\endtabular}
}

\newlength\cmsTabSkip\setlength{\cmsTabSkip}{1ex}

\newcommand{\sqrtthirteentev}{\ensuremath{\sqrt{s}=13\TeV}\xspace}
\newcommand{\uubar}{\ensuremath{\PQu\PAQu}\xspace}
\newcommand{\ddbar}{\ensuremath{\PQd\PAQd}\xspace}
\newcommand{\ssbar}{\ensuremath{\PQs\PAQs}\xspace}
\newcommand{\zjets}{\ensuremath{\PZ+\text{jets}}\xspace}
\newcommand{\Zlljets}{\ensuremath{\PZ(\Pell\Pell)+\text{jets}}\xspace}
\newcommand{\wjets}{\ensuremath{\PW+\text{jets}}\xspace}
\newcommand{\phojets}{\ensuremath{\PGg+\text{jets}}\xspace}
\newcommand{\Zvvjets}{\ensuremath{\PZ(\PGn\PGn)+\text{jets}}}
\newcommand{\Wlvjets}{\ensuremath{\PW(\Pell\PGn)+\text{jets}}\xspace}

\newcommand{\mZpr}{\ensuremath{m_{\PZpr}}\xspace}
\newcommand{\mdm}{\ensuremath{m_{\mathrm{DM}}}\xspace}
\newcommand{\mmed}{\ensuremath{m_{\mathrm{med}}}\xspace}
\newcommand{\gq}{\ensuremath{g_{\PQq}}\xspace}
\newcommand{\gdm}{\ensuremath{g_{\mathrm{DM}}}\xspace}
\providecommand{\PWst}{{\HepParticle{\PW}{}{\ast}}\xspace}

\cmsNoteHeader{SUS-23-017}
\title{Search for dark matter produced in association with a low-multiplicity jet in proton-proton collisions at \texorpdfstring{$\sqrt{s} = 13\TeV$}{sqrt(s) = 13 TeV}}

\date{\today}

\abstract{
A search for dark matter particles is presented using events with large missing transverse momentum and a single energetic jet containing few constituents. The analysis uses proton-proton collision data recorded by the CMS experiment at a center-of-mass energy of 13\TeV. This corresponds to an integrated luminosity of 138\fbinv, recorded during the 2016--2018 LHC data-taking period. This is the first search at the LHC to employ the low-multiplicity jet signature. Supervised machine learning is used to maximize sensitivity to potential signals in this challenging final state. No excess of events over the standard model background prediction is observed. The results are used to set upper limits on the dark matter production cross section at 95\% confidence level (\CL) within simplified models featuring vector and axial-vector mediators. For a dark matter particle mass of approximately 150 (550)\GeV, mediator masses up to 4250 (3500)\GeV are excluded at 95\% \CL.
}

\hypersetup{
pdfauthor={CMS Collaboration},
pdftitle={Search for dark matter recoiling from a low-multiplicity jet at 13 TeV},
pdfsubject={CMS},
pdfkeywords={CMS, dark matter, machine learning, PencilJets, MonoJet}}

\maketitle

\section{Introduction}
\label{sec:intro}

The standard model (SM) of particle physics is a remarkably successful framework, with predictions confirmed by a wide range of experimental measurements. Despite this success, it does not account for several fundamental aspects of the universe, such as gravity and the existence of dark matter (DM). Although compelling astrophysical and cosmological evidence supports the presence of DM~\cite{Feng:2010gw,Porter:2011nv,Clowe:2003tk,Planck:2015fie}, its particle nature remains unknown, and no direct evidence has yet been observed. To address these gaps, it is essential to explore potential evidence for physics beyond the standard model (BSM).  
 
Searches for BSM are commonly interpreted within three complementary theoretical frameworks: effective field theories (EFTs)~\cite{Willenbrock_2014}, simplified models~\cite{Abercrombie:2015wmb, Busoni:2013lha, Buchmueller:2014yoa}, and ultraviolet (UV) complete theories~\cite{Allanach:2702850}. These approaches differ in their levels of complexity and predictive power. EFTs provide a model-independent description, valid at low energies but may break down near the energy scales probed at the LHC. UV-complete models offer a fully consistent description at all energy scales but often involve a large number of parameters. Simplified models provide an intermediate approach by introducing a minimal set of new particles and interactions that capture the essential phenomenology relevant for collider searches; as such, they serve as a practical bridge between EFTs and UV-complete theories.

\begin{figure}[htb!]
   \centering
      \includegraphics[width=0.6\columnwidth]{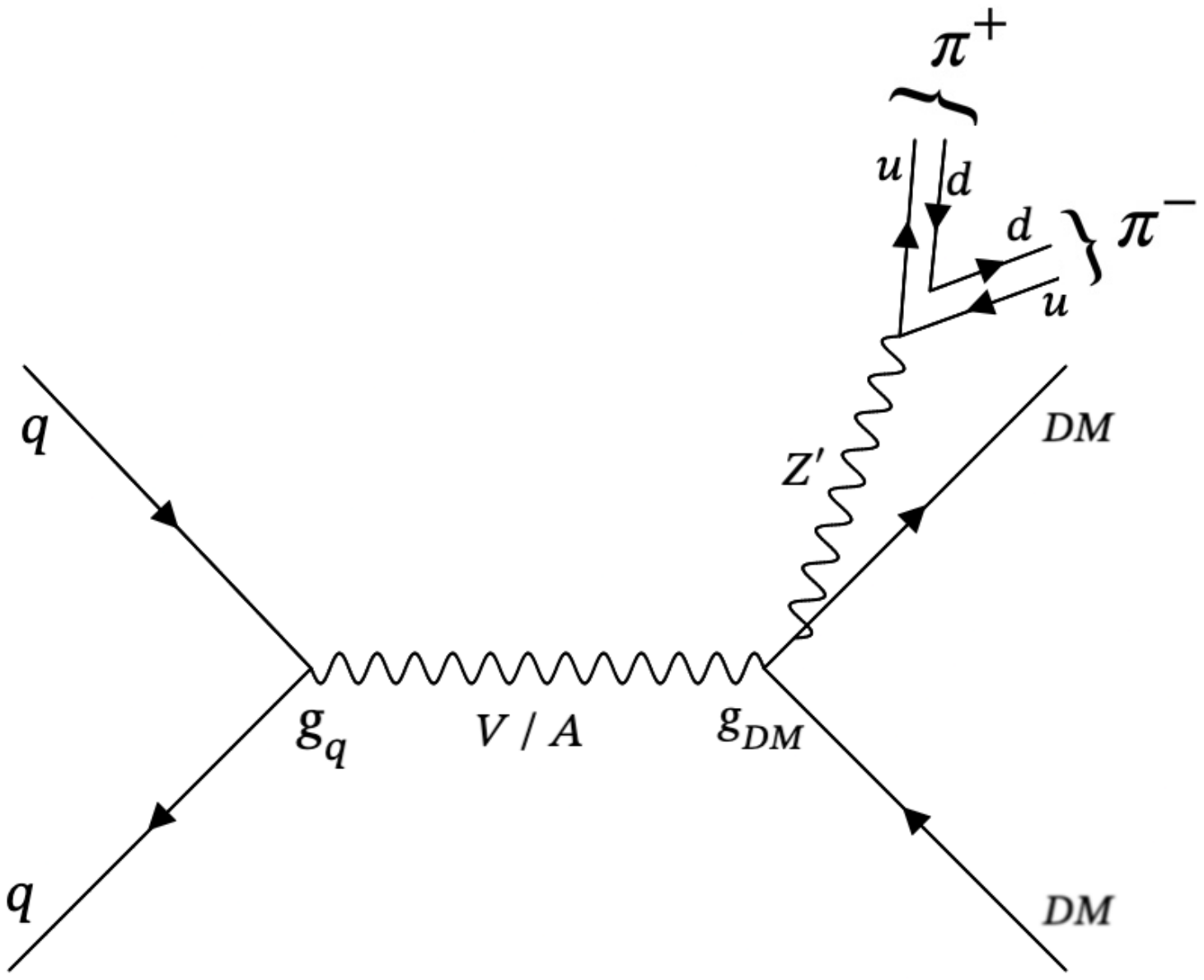}
      \caption{Representative leading order Feynman diagram of the signal process, where DM is the dark matter candidate, V/A represents the vector or axial-vector mediator, $\gq$ and $\gdm$ represents the mediator coupling to the quarks and DM respectively. Due to the low mass of the \PZpr boson (${\approx}1$\GeV), it primarily decays to a low-multiplicity final state, typically two pions, and is expected to be reconstructed in the detector as a jet with very narrow cone of radiation, called the pencil-jet.}
      \label{fig:feynmandiag}
\end{figure}

This paper considers a simplified model featuring an Abelian dark sector with a light \GeV-scale \PZpr boson. This topology offers a distinctive signature of dark-sector dynamics that can significantly extend CMS sensitivity beyond conventional mono-jet searches. The primary process involves the production of a pair of DM particles, with one of the DM particles radiating a \PZpr boson through final-state radiation (FSR), as illustrated in Fig.~\ref{fig:feynmandiag}. The \PZpr boson subsequently decays into SM particles, producing a visible experimental signature, while the DM particles escape detection, resulting in significant missing transverse momentum. Two signal models are considered when interpreting the results~\cite{MonoZprime_Theory}. In one scenario, the mediator couples to both quarks, and DM, through vector interactions, while in the other, it couples through axial-vector interactions.

The model assumes a leptophobic \PZpr boson that predominantly decays into quarks. Decays to leptons are not considered, as a \PZpr boson in the mass range explored here would likely have already been observed in previous experiments. The analysis probes DM candidate masses in the range 0.1--1\TeV, mediator masses between 2 and 5\TeV, and a light \PZpr boson with a mass between 0.3 and 3\GeV. These choices span a broad class of weakly interacting massive particle (WIMP)~\cite{Jungman1996} scenarios, which remain well-motivated in many extensions of the SM~\cite{Roszkowski_2018} and ensure sensitivity to resonant production while remaining consistent with existing constraints.

Unlike traditional mono-X searches, where a single detectable SM particle (denoted as X) recoils against missing transverse momentum, the signal considered here receives contributions from both initial-state radiation (ISR) and FSR as described in Ref.~\cite{MonoZprime_Theory}. In this model, FSR contributions can be comparable to, or even exceed, ISR contributions, depending on the parameter space. When the \PZpr boson mass ($\mZpr$) exceeds twice the DM mass (\mdm), it predominantly decays into DM particles. This scenario can be effectively probed through a monojet signature, as previously studied by the ATLAS and CMS collaborations~\cite{CMS:2021far, CMS:2017zts, ATLAS:2018nda, ATLAS:2021kxv}. For $\mZpr \leq 2\mdm$, the \PZpr boson cannot decay directly into DM and instead decays to SM particles through higher-dimensional operators. In this search, $\mZpr$ is restricted to values below 3\GeV to avoid crossing the charm threshold, above which the increased hadronic multiplicity would obscure the characteristic narrow-jet signature targeted by this analysis. Owing to kinematic constraints, a \GeV-scale \PZpr boson decays primarily into light quark pairs (\uubar, \ddbar, \ssbar).

Due to its small mass and large Lorentz boost, the decay products of the \PZpr boson are highly collimated, forming a narrow cone of radiation with a characteristic angular size proportional to ${\mZpr}/{\pt(\PZpr)}$, where $\pt(\PZpr)$ is the transverse momentum of the \PZpr boson. 
As a result, the decay is typically reconstructed as a single jet with low charged-particle multiplicity and, in many cases, a distinctive two-prong substructure (where a ``prong" corresponds to a charged track), as illustrated in Fig.~\ref{fig:feynmandiag}. Such jets, referred to as ``pencil-jets" in the analysis, exhibit features closely resembling those of hadronically decaying \PGt leptons.
This similarity is exploited in the analysis by employing the hadrons-plus-strips algorithm~\cite{Sirunyan_2018}, originally developed for the reconstruction of hadronic \PGt decays in the CMS detector. In particular, both boosted \PZpr jets and hadronic \PGt decays are characterized by a narrow radiation profile and low particle multiplicity, providing strong discrimination against generic quantum chromodynamics (QCD) jets. 

The paper is organized as follows. Section~\ref{sec:detector} introduces the CMS detector and the reconstruction of the physics objects and event variables used in the analysis. Section~\ref{sec:samples} describes the dataset and the Monte Carlo (MC) simulated samples. The event selection is presented in Section~\ref{sec:selection}, while Section~\ref{sec:ml_model} outlines the signal extraction strategy. Systematic uncertainties are discussed in Section~\ref{sec:uncertainty}. The results and their interpretation are presented in Section~\ref{sec:combine}, and a summary of the analysis is given in Section~\ref{sec:summary}. The tabulated results are also provided in the HEPData record~\cite{hepdata}.

\section{The CMS detector and event reconstruction}
\label{sec:detector}
The CMS apparatus~\cite{Chatrchyan:2008zzk,CMS:2023gfb} is a multipurpose, nearly hermetic detector, designed to trigger on~\cite{Sirunyan:2020zal,Khachatryan:2016bia,CMS:2024aqx} and identify electrons, muons, photons, and (charged and neutral) hadrons~\cite{CMS:2020uim,muon_cms,CMS:2014pgm}. Its central feature is a superconducting solenoid of 6\unit{m} internal diameter, providing a magnetic field of 3.8\unit{T}. Within the solenoid volume are a silicon pixel and strip tracker, a lead tungstate crystal electromagnetic calorimeter (ECAL), and a brass and scintillator hadron calorimeter (HCAL), each composed of a barrel and two endcap sections. Forward calorimeters extend the pseudorapidity coverage provided by the barrel and endcap detectors. Muons are reconstructed using gas-ionization detectors interleaved with the layers of the steel flux-return yoke outside the solenoid. 

The silicon tracker used in 2016 measured charged particles within the range $\abs{\eta} < 2.5$. For non-isolated particles of $1 < \pt < 10\GeV$ and $\abs{\eta} < 1.4$, the track resolutions were typically 1.5\% in \pt and 25--90 (45--150)\mum in the transverse (longitudinal) impact parameter~\cite{CMS:2014pgm}. At the start of 2017, a new pixel detector was installed~\cite{Phase1Pixel}; the upgraded tracker measured particles up to $\abs{\eta} = 3.0$ with typical resolutions of 1.5\% in \pt and 20--75\mum in the transverse impact parameter~\cite{DP-2020-049} for non-isolated particles of $1 < \pt < 10\GeV$. According to simulation studies~\cite{DP-2017-015}, similar improvements are expected in the longitudinal direction. 

The electron momentum is estimated by combining the energy measurement in the ECAL with the momentum measurement in the tracker. The momentum resolution for electrons with $\pt \approx 45\GeV$ from $\PZ \to \Pe \Pe$ decays ranges 1.6--5.0\%. It is generally better in the barrel region than in the endcaps, and also depends on the bremsstrahlung energy emitted by the electron as it traverses the material in front of the ECAL~\cite{CMS:2020uim,CMS-DP-2020-021}. When combining information from the entire detector, the jet energy resolution typically amounts to 15--20\% at 30\GeV, 10\% at 100\GeV, and 5\% at 1\TeV~\cite{Khachatryan:2016kdb}. 

In the region $\abs{\eta} < 1.74$, the HCAL cells have widths of 0.087 in pseudorapidity and 0.087 in azimuth ($\phi$). In the $\eta$-$\phi$ plane, and for $\abs{\eta} < 1.48$, the HCAL cells map on to $5 \times 5$ arrays of ECAL crystals to form calorimeter towers projecting radially outwards from close to the nominal interaction point. For $\abs{\eta} > 1.74$, the coverage of the towers increases progressively to a maximum of 0.174 in $\Delta \eta$ and $\Delta \phi$. Within each tower, the energy deposits in ECAL and HCAL cells are summed to define the calorimeter tower energies, which are subsequently used to provide the energies and directions of hadronic jets. The hadron forward (HF) calorimeter uses steel as an absorber and quartz fibers as the sensitive material. The two halves of the HF are located 11.2\unit{m} from the interaction region, one on each end, and together they provide coverage in the range $3.0 < \abs{\eta} < 5.2$. 

Events of interest are selected using a two-tiered trigger system. The first level trigger, composed of custom hardware processors, uses information from the calorimeters and muon detectors to select events at a rate of around 100\unit{kHz} within a fixed latency of about 4\mus~\cite{Sirunyan:2020zal}. The second level, known as the high-level trigger, consists of a farm of processors running a version of the full event reconstruction software optimized for fast processing, and reduces the event rate to a few\unit{kHz} before data storage~\cite{Khachatryan:2016bia,CMS:2024aqx}. A more detailed description of the CMS detector, together with a definition of the coordinate system used and the relevant kinematic variables, can be found in Ref.~\cite{Chatrchyan:2008zzk}.

The primary vertex is taken to be the vertex corresponding to the hardest scattering in the event, evaluated using tracking information alone, as described in Section 9.4.1 of Ref.~\cite{CMS-TDR-15-02}.
The global event reconstruction, also known as particle-flow (PF) event reconstruction~\cite{CMS-PRF-14-001}, aims to reconstruct and identify each individual particle in an event, with an optimized combination of all subdetector information. In this process, the identification of the particle type (photon, electron, muon, charged hadron, neutral hadron) plays an important role in the determination of the particle direction and energy. Photons (\eg coming from \PGpz decays or from electron bremsstrahlung) are identified as ECAL energy clusters not linked to the extrapolation of any charged particle trajectory to the ECAL. Electrons (\eg coming from photon conversions in the tracker material or from \PB hadron semileptonic decays) are identified as a primary charged particle track and potentially many ECAL energy clusters corresponding to this track extrapolation to the ECAL and to possible bremsstrahlung photons emitted along the way through the tracker material. Muons (\eg from \PB hadron semileptonic decays) are identified as tracks in the central tracker consistent with either a track or several hits in the muon system, and associated with calorimeter deposits compatible with the muon hypothesis. Charged hadrons are identified as charged particle tracks neither identified as electrons, nor as muons. Finally, neutral hadrons are identified as HCAL energy clusters not linked to any charged hadron trajectory, or as a combined ECAL and HCAL energy excess with respect to the expected charged hadron energy deposit.

For each event, hadronic jets are clustered from these reconstructed particles using the infrared- and collinear-safe anti-\kt algorithm~\cite{Cacciari:2008gp, Cacciari:2011ma} with a distance parameter ($\Delta R$) of 0.4, where $\Delta R = \sqrt{\smash[b]{(\Delta \phi)^2 + (\Delta \eta)^2}}$. These jets are referred to as ``AK4 jets". Jet momentum is determined as the vectorial sum of all particle momenta in the jet, and is found from simulation to be, on average within 5--10\% of the true momentum over the whole \pt spectrum and detector acceptance. Additional $\Pp\Pp$ interactions within the same or nearby bunch crossings (pileup) can contribute extra tracks and calorimetric energy depositions to the jet momentum. To mitigate this effect, charged particles identified to be originating from pileup vertices are discarded and an offset correction is applied to correct for remaining contributions~\cite{CMS:2020ebo}. Jet energy corrections are derived from simulation to bring the measured response of jets to that of particle level jets on average. In situ measurements of the momentum balance in dijet, $\PGg + \text{jet}$, $\PZ + \text{jet}$, and multijet events are used to account for any residual differences in the jet energy scale (JES) between data and simulation~\cite{Khachatryan:2016kdb}. The jet energy resolution (JER) amounts typically to 15--20\% at 30\GeV, 10\% at 100\GeV, and 5\% at 1\TeV~\cite{Khachatryan:2016kdb}. Additional selection criteria are applied to each jet to remove jets potentially dominated by anomalous contributions from various subdetector components or reconstruction failures. 

To target the characteristic \PZpr boson signature, the hadrons-plus-strips algorithm~\cite{Sirunyan_2018} is employed. This algorithm reconstructs narrow jets by combining charged-particle tracks with calorimeter energy deposits. In this paper, such narrow jets are referred to as pencil-jets. A pencil-jet is defined as a reconstructed object containing at most three tracks, charged hadrons, and electromagnetic clusters within a cone of size $\Delta R < 0.1$, with the exact cone size depending on the $\pt$ of the hadronic system~\cite{Sirunyan_2018}. In addition, the pencil-jet is required to be isolated from other particle-flow (PF) candidates within a cone of $\Delta R = 0.5$. The leading AK4 jet reconstructed in the same $\eta$-$\phi$ region as the pencil-jet is referred to as the associated jet. To further suppress backgrounds from QCD jets, electrons, and muons, the \textsc{DeepTau} algorithm~\cite{CMS:2022prd,Sirunyan_2018} is applied. This choice is motivated by the similarity between pencil-jets and hadronically decaying $\Pgt$ leptons, as discussed in later sections. Only candidates with a high \textsc{DeepTau} score are retained. Separate discriminants are used to reject QCD jets, electrons, and muons, corresponding to jet identification efficiencies of approximately 90\%, 80\%, and 99.5\%, respectively, and misidentification rates of about 2\%, 0.1\%, and 0.02\%. These efficiency and misidentification rate working points were selected based on a tradeoff between maximizing signal efficiency and ensuring sufficient background contribution for robust systematic estimation.

Electrons (photons) are required to have $\pt > 10 (15)\GeV$ and $\abs{\eta} < 2.5$, and must satisfy 'loose' identification and isolation criteria corresponding to approximately 90\% reconstruction efficiency~\cite{CMS:2020uim}. Muons are required to have $\pt > 10\GeV$ and $\abs{\eta} < 2.4$, and must satisfy 'loose' identification criteria corresponding to a 90\% efficiency~\cite{muon_cms}. The AK4 jets that are identified as originating from bottom quarks are classified as ``b-tagged jets". These jets must satisfy $\pt > 20\GeV$ and $\abs{\eta} < 2.4$, and are identified using the \textsc{DeepCSV} algorithm~\cite{Sirunyan:2017ezt} with the ``medium" working point. This working point corresponds to a b-jet identification efficiency of approximately 80\% and a misidentification probability of about 10\% for jets originating from light-flavor quarks or gluons.

The missing transverse momentum vector \ptvecmiss is computed as the negative vector sum of the transverse momenta of all the PF candidates in an event, and its magnitude is denoted as PF \ptmiss~\cite{Sirunyan:2019kia}. The \ptvecmiss is modified to account for corrections to the energy scale and resolution of the reconstructed jets in the event.  Anomalous high-\ptmiss events can be due to reconstruction failures, detector malfunctions, or noncollisional backgrounds. Such events are rejected by dedicated filters that are designed to eliminate more than 85--90\% of the spurious high-\ptmiss events with a signal efficiency exceeding 99.9\%~\cite{Sirunyan:2019kia}.

\section{Signal and background events simulation} 
\label{sec:samples}
The analysis presented in this paper uses data collected by the CMS experiment in 2016, 2017, and 2018 at \sqrtthirteentev, corresponding to integrated luminosities of 36.3, 41.5, and 59.7\fbinv, respectively. The 2016 data set is further divided into two sub-periods, 2016-preVFP and 2016-postVFP eras, corresponding to integrated luminosities of 16.1 and 20.2\fbinv, respectively, to account for changes in the CMS detector conditions following an intervention on the pixel detector during data taking.

All background contributions are estimated from simulated samples generated with MC event generators. The corresponding production cross sections and references are summarized in Table~\ref{tab:BackgroundSamples}.
The dominant background arises from $\Zvvjets$ production, where the \PZ boson decays invisibly to neutrinos, producing genuine missing transverse momentum. The second-largest background originates from \Wlvjets production, when the \PW boson decays leptonically and the charged lepton is not reconstructed, falls outside the detector acceptance, or fails the selection criteria. Background contributions from \Zlljets events arise when the charged leptons from the \PZ boson decay are not reconstructed or fall outside the detector acceptance, allowing the event to mimic the signal signature. 
The contribution from \phojets events in the signal region is negligible and is therefore combined with the \Zlljets background in the signal region distributions and 
Table~\ref{tab:BackgroundSamples}.
Top quark processes, including \ttbar and single top production, also contribute to the background when the \PW boson produced in the top quark decay subsequently decays leptonically, resulting in genuine missing transverse momentum. The \Zvvjets, \Wlvjets, \Zlljets, \phojets and \ttbar samples are generated at next-to-leading order (NLO) in QCD using the \MGvATNLO generator, while single top processes are generated at NLO with the \POWHEG generator.
Diboson production ($\PW\PW$, $\PW\PZ$, and $\PZ\PZ$) constitutes an additional background source when one of the bosons decays leptonically, producing missing transverse momentum, while the other boson decays hadronically. These samples are generated at leading-order (LO) using \PYTHIAviii. Finally, QCD multijet events can contribute due to jet energy mismeasurement that leads to large apparent missing transverse momentum. Although such events typically do not contain genuine missing transverse momentum, their large production cross section results in a non-negligible contribution. These samples are generated at LO in QCD using the \MGvATNLO generator in bins of hadronic transverse momentum.

{\tolerance=1200
The propagation of all final-state particles through the CMS detector is simulated using the \GEANTfour-based CMS detector simulation framework~\cite{Agostinelli:2002hh}, and the same event reconstruction algorithms as those used for data are applied to the simulated samples. Simulated events also include the effects of additional $\Pp\Pp$ interactions (pileup), with the distribution of reconstructed pileup vertices reweighted to match that observed in data. The average number of pileup interactions per bunch crossing is approximately 23, 32, and 33 for the 2016~\cite{CMS:2021xjt}, 2017~\cite{CMS-PAS-LUM-17-004}, and 2018~\cite{CMS-PAS-LUM-18-002} data-taking periods used in this analysis, respectively, assuming a total inelastic $\Pp\Pp$ cross section of 69.2\unit{mb}~\cite{CMS:2021xjt,CMS-PAS-LUM-17-004,CMS-PAS-LUM-18-002}.
\par}

Signal MC events are generated using the simplified model described in Ref.~\cite{MonoZprime_Theory}, following the recommendations of the ATLAS-CMS Dark Matter Forum~\cite{Abercrombie:2015wmb}. A representative Feynman diagram for this model is shown in Fig.~\ref{fig:feynmandiag}. In this scenario, the \PZpr boson is assumed to decay into a low-multiplicity hadronic final state, typically consisting of two pions. This decay is modeled using the $\PZpr \to \uubar$ process, following Ref.~\cite{MonoZprime_Theory}. For \PZpr masses above $2m_{\PK}$, where $m_{\PK}$ denotes the kaon mass, decays to $s\bar{s}$ are also simulated. Studies show that the detector response within the kinematic range of the search region is similar for \PZpr decays to pion or kaon pairs. The coupling of the mediator (both vector and axial vector) to SM quarks (\gq) is set to 0.25, while the coupling to dark matter candidates (\gdm) is set to 1.0. Signal samples are generated for different hypotheses by varying the dark matter mass between 0.1 and 1\TeV and the mediator mass between 2 and 5\TeV.

The production cross sections remain consistent within theoretical uncertainties when the \PZpr mass is varied between 0.3 and 3\GeV, and studies across this range yield similar overall results. Alternative decay modes, such as $\PZpr \to \PQs\PAQs$, were also examined, with kinematic distributions and exclusion limits found to be consistent within uncertainties. The upper bound of $m_{\PZpr}=3\GeV$ is chosen to avoid crossing the charm threshold, as discussed in Section~\ref{sec:intro}. For lower mediator masses, particularly when $m_{\PZpr} < 2m_{\PGp}$, the cross sections decrease significantly because of the reduced phase space available for decays to SM particles, that lowers the probability of producing detectable signatures.

\section{Event selection}
\label{sec:selection}
Events are selected using high-level trigger paths that require thresholds of 120\GeV on both the PF missing transverse momentum, \ptmiss, and the hadronic missing transverse momentum, \mht. The PF \ptmiss is calculated from PF candidates reconstructed at the trigger level, while \mht is defined as the magnitude of the vector \ptvec sum of jets with $\pt > 20\GeV$ and $\abs{\eta} < 5.0$, also reconstructed at the trigger level. To suppress anomalous events originating from detector noise, the neutral hadron energy fraction of these jets is required to be less than 90\%. For the trigger selections used in this analysis, muon candidates are not included in the PF \ptmiss or \mht calculations. This choice is motivated by the expected final-state signature of the signal. The trigger efficiency is measured to be $96\%$ for events passing the analysis selection for $\ptmiss > 250\GeV$ and becomes more than 99\% efficient for events with $\ptmiss > 350\GeV$.

Events containing electrons, muons, or photons that satisfy the ``loose'' object identification criteria described in Section~\ref{sec:detector} are vetoed. To further suppress backgrounds from top quark production, events containing a b-tagged jet are rejected according to the previously defined b-tagging criteria. Additional requirements are applied to reduce events with anomalous \ptmiss arising from reconstruction failures of the PF algorithm. In particular, events must satisfy ((PF \ptmiss - calorimeter \ptmiss)/PF \ptmiss$)<0.5$ where calorimeter \ptmiss is the missing transverse momentum, computed using only calorimeter information. In this paper, PF \ptmiss is also referred to as the hadronic recoil. Candidate events are required to have a hadronic recoil of at least $250\GeV$.

Events are required to have at least one pencil-jet with $\pt>140\GeV$ and $\abs{\eta}<2.3$. The associated jet must satisfy quality requirements based on the composition of the jet in terms of PF candidates. Specifically, the charged hadron energy fraction is required to be greater than 10\%, and the neutral hadron energy fraction must be less than 80\%~\cite{CMS-PAS-JME-16-003}. These criteria help suppress events arising from beam-induced backgrounds and calorimeter noise. In addition, the associated jet is required to be the highest $\pt$ AK4 jet in the event, a condition that is satisfied in approximately 95\% of signal events according to simulation. The contribution from QCD multijet events, where large PF \ptmiss can arise from jet momentum mismeasurement, is further suppressed by requiring the azimuthal separation between the \ptvecmiss direction and the leading pencil-jet to be greater than 0.5.

To bring the simulation into better agreement with experimental data, event weights are applied to MC simulated events. These weights incorporate data-driven scale factors (SFs), derived from orthogonal data samples, to correct for differences in trigger performance, object reconstruction, and identification efficiencies between data and simulation. Additional corrections are also applied to account for higher-order theoretical effects. In particular, per-event weights are used to include higher-order QCD and electroweak (EW) corrections and their associated uncertainties, following the prescription described in Ref.~\cite{DMTheory}.

\begin{figure*}[hbtp]
   \centering
      \includegraphics[width=0.49\textwidth]{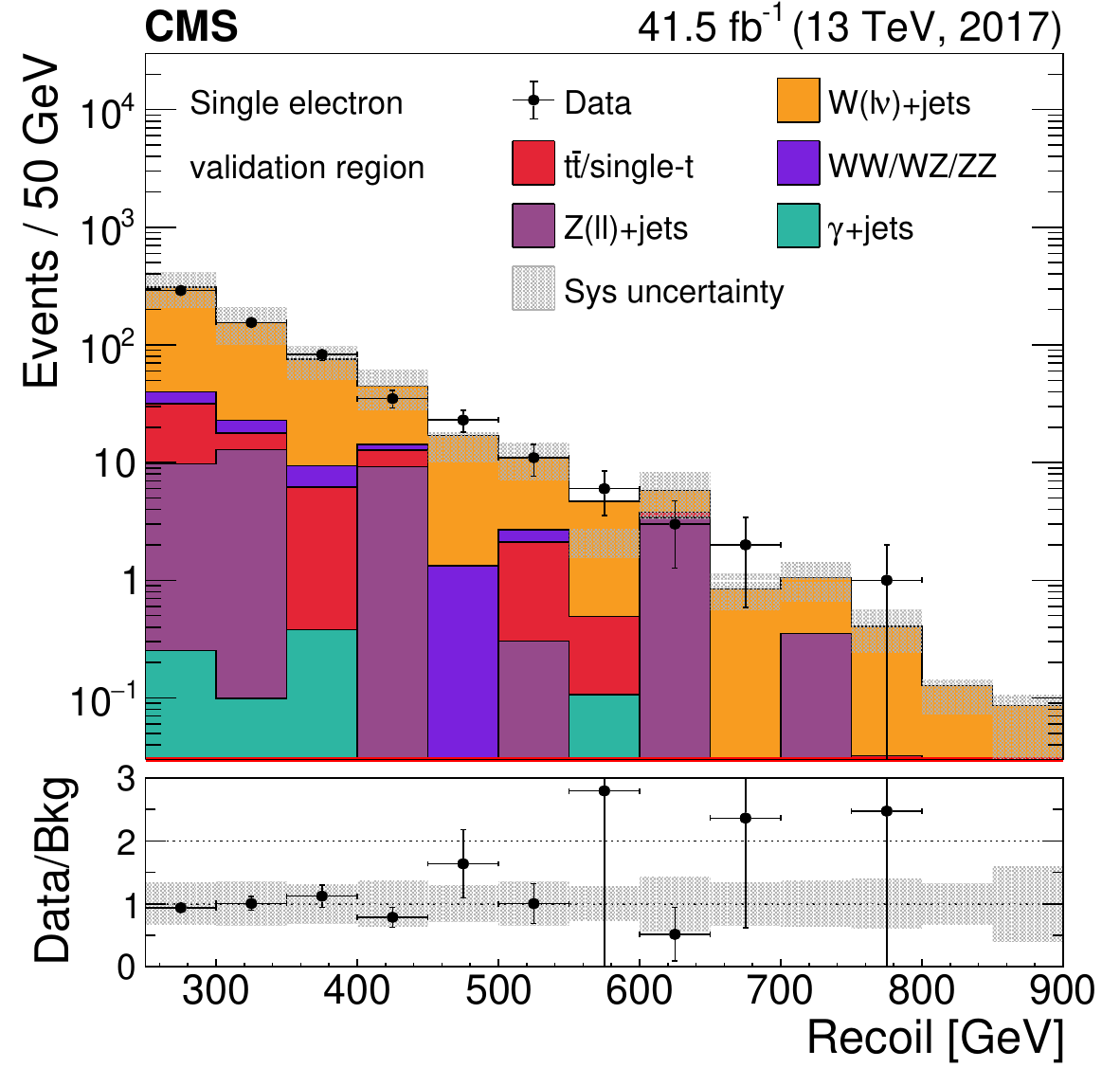}
      \includegraphics[width=0.49\textwidth]{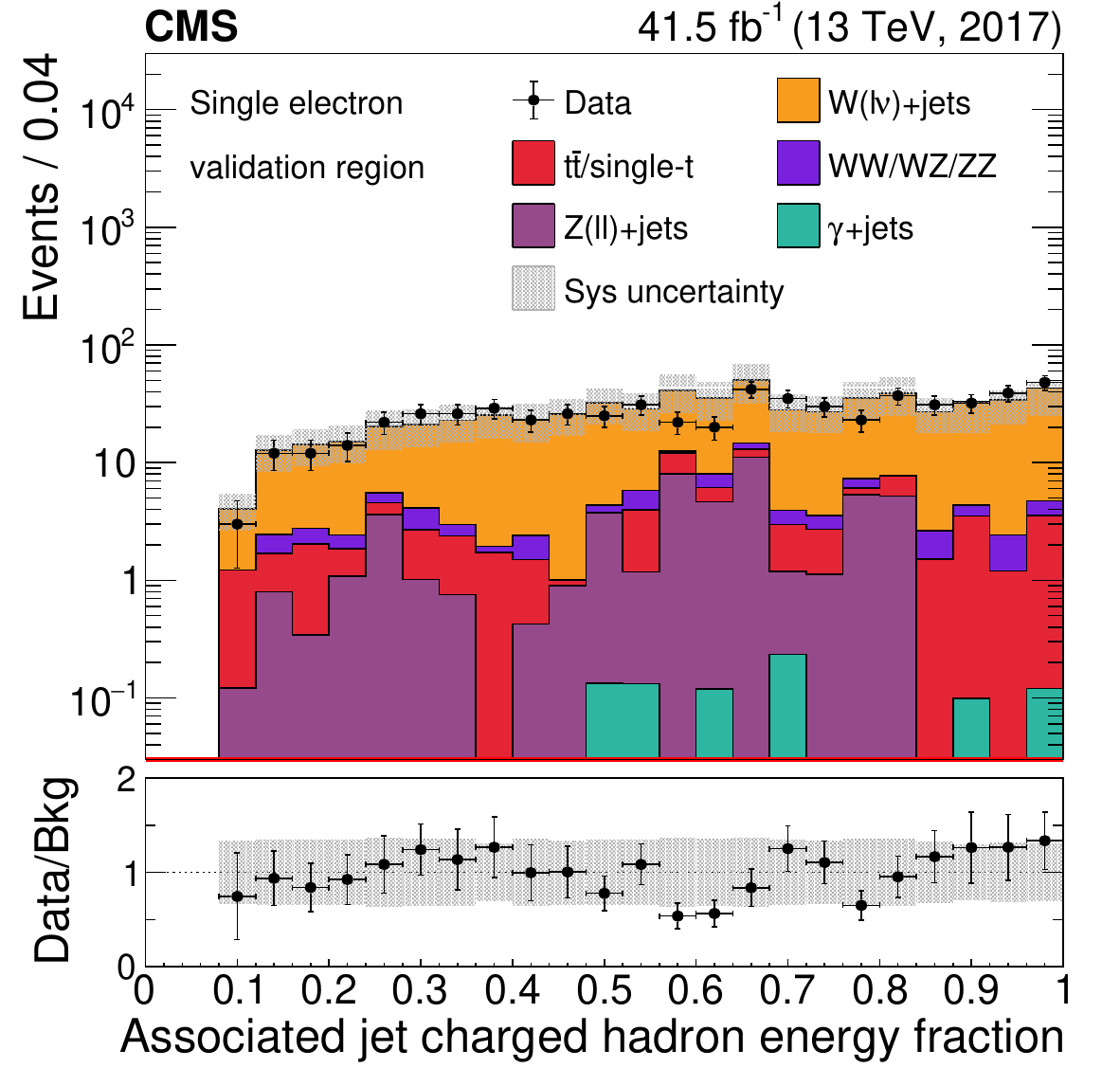}\\
      \includegraphics[width=0.49\textwidth]{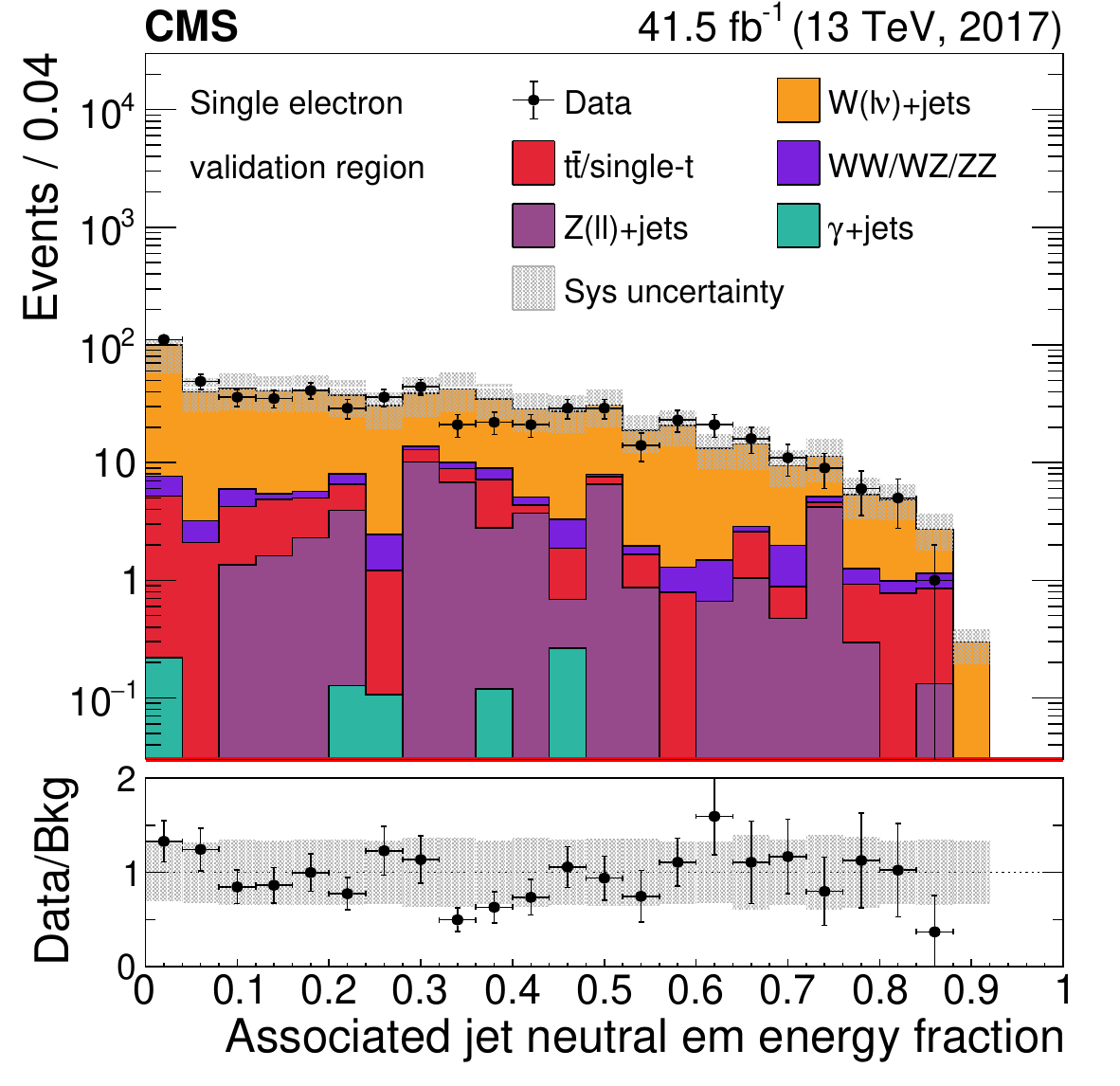}
      \includegraphics[width=0.49\textwidth]{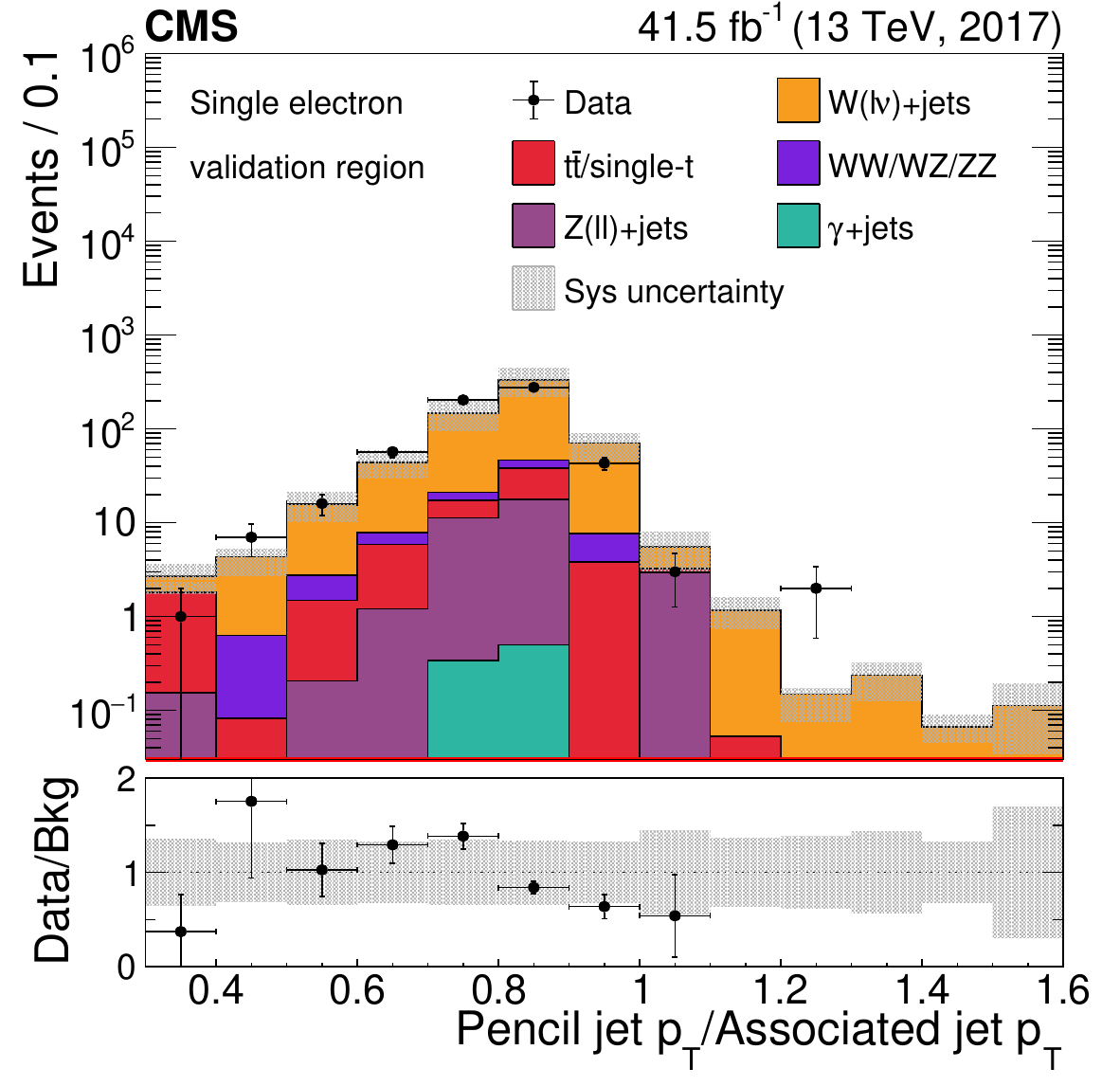}
      \caption{Comparison between observed data and simulated backgrounds in the single electron validation region for the recoil (upper-left), the charged hadron energy fraction of the associated jet (upper-right), the neutral electromagnetic energy fraction of the associated jet (lower-left), and the ratio of pencil-jet \pt to the associated jet \pt (lower-right). This region is orthogonal to the signal region, resulting in a negligible expected signal contribution. Total  systematic uncertainties are labeled as ``Sys uncertainity" in the legend box. The ratio of data to the pre-fit background prediction is shown in the bottom pads, with the grey band indicating the total systematic uncertainty.}
      \label{fig:single_ele_cr_2017}
\end{figure*}  

Because there is no well-established SM object that directly reproduces the two-prong jet signature targeted in this analysis, hadronic decays of the \PGt lepton are used to derive the pencil-jet scale factors and their associated uncertainty. At high \pt, the decay products of three-prong \PGt decays can become highly collimated, causing them to appear as two-prong structures in the detector. Similarly, one-prong \PGt decays may be reconstructed as two-prong objects if an additional nearby track is mistakenly associated with the decay products. These features of hadronic \PGt decays therefore provide a suitable proxy for calibrating the pencil-jet reconstruction performance and for deriving the corresponding scale factors and uncertainties. 

The pencil-jet scale factors and uncertainties are extracted from a simultaneous fit that accounts for contributions from jets misidentified as \PGt leptons. The fit is performed on transverse mass distributions in two complementary samples: a $\PWst\to\PGt\PGn$ sample constructed using the pencil jet and PF \ptmiss, and a $\PWst\to\PGm\PGn$ sample constructed using the muon and PF \ptmiss. To ensure statistical independence from the main analysis, the procedure is carried out in a dedicated region with only about 10\% overlap with the signal region. After applying these corrections, the simulation is validated using dedicated validation regions.

\begin{table*}[htb]
\centering
\topcaption{Simulated datasets used to model the background processes. The table summarizes the expected yields of each background contribution in the signal region for the corresponding data-taking years, as obtained from the background-only fit. Total systematic uncertainties are reported including both upward and downward variations. The production cross sections for each process, along with their references, are also provided.}
\renewcommand{\arraystretch}{1.0}
\begin{scotch}{lccccc}
    Backgrounds          & 2016 yields   & 2017 yields   & 2018 yields   & {Cross section (pb)} & References \\
    \hline
    \Zvvjets             & $900\pm 80$   & $1100\pm 100$ & $1300\pm 110$ &    9.9e+02           & ~\cite{DMTheory}\\
    \Wlvjets             & $580\pm 90$   & $680\pm 90$   & $830\pm 120$  &    7.2e+02           & ~\cite{Alwall_2014}\\
    QCD multijets         & $260\pm 110$  & $330\pm 140$  & $370\pm 180$  &    2.2e+08           & ~\cite{CMS:2021far}\\
    $\PW\PW/\PW\PZ/\PZ\PZ$   & $48\pm 12$    & $63\pm 15$    & $72\pm 18$    &    1.9e+02           & ~\cite{WW_xsec,WZ_xsec,ZZ_xsec}\\
    \ttbar/single-\PQt         & $40\pm 10$    & $31\pm 6$     & $40\pm 10$    &    1.1e+03       & ~\cite{ttbar_xsec,ST_tchannel_xsec,ST_tW_xsec}\\
    \Zlljets             & $25\pm 6$     & $33\pm 8$     & $40\pm 12$    &    6.3e+03           & ~\cite{Alwall_2014}\\
\end{scotch}
\label{tab:BackgroundSamples}
\end{table*}

\begin{figure*}[hbtp]
   \centering
      \includegraphics[width=0.49\textwidth]{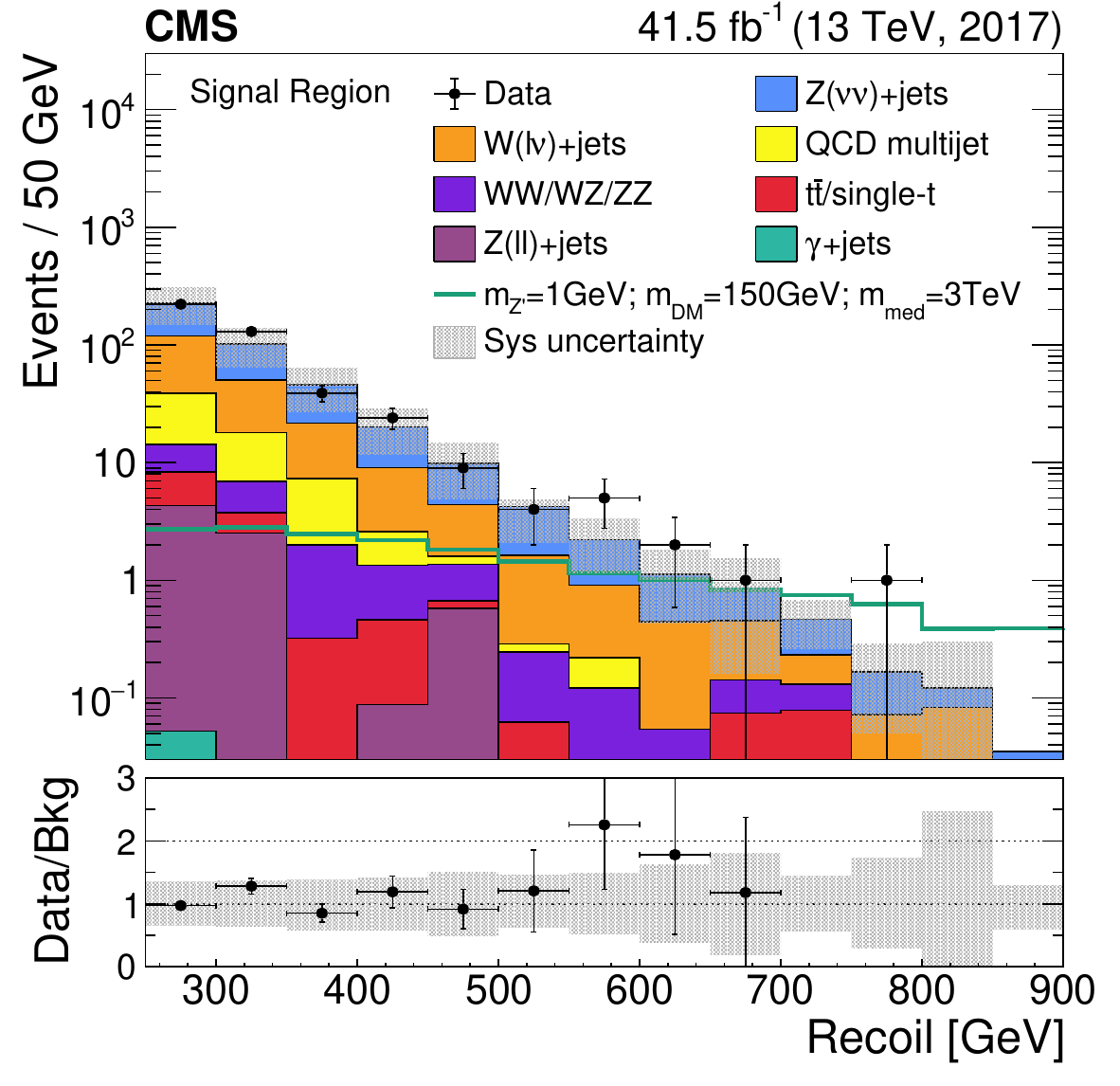}
      \includegraphics[width=0.49\textwidth]{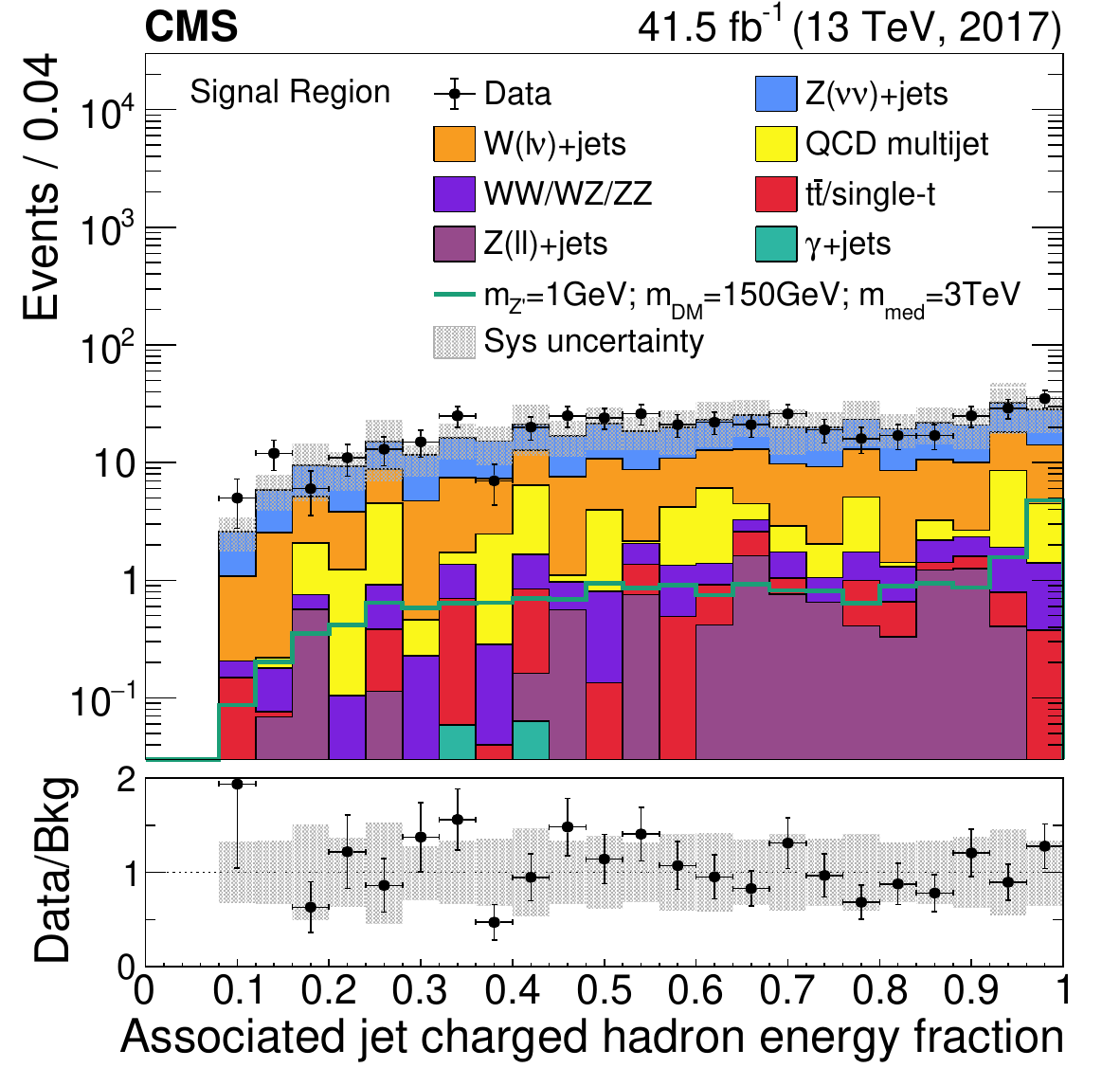}\\
      \includegraphics[width=0.49\textwidth]{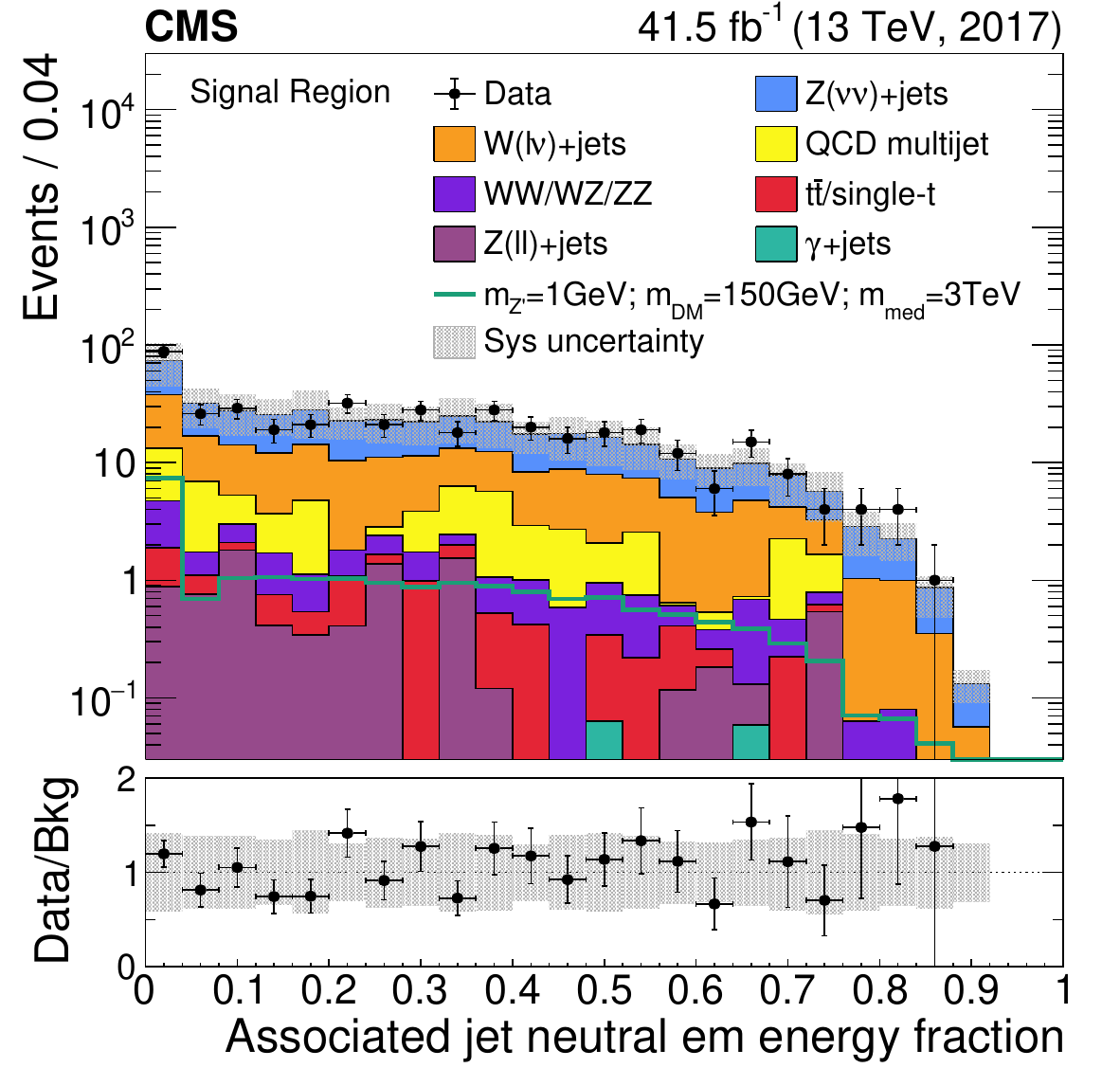}
      \includegraphics[width=0.49\textwidth]{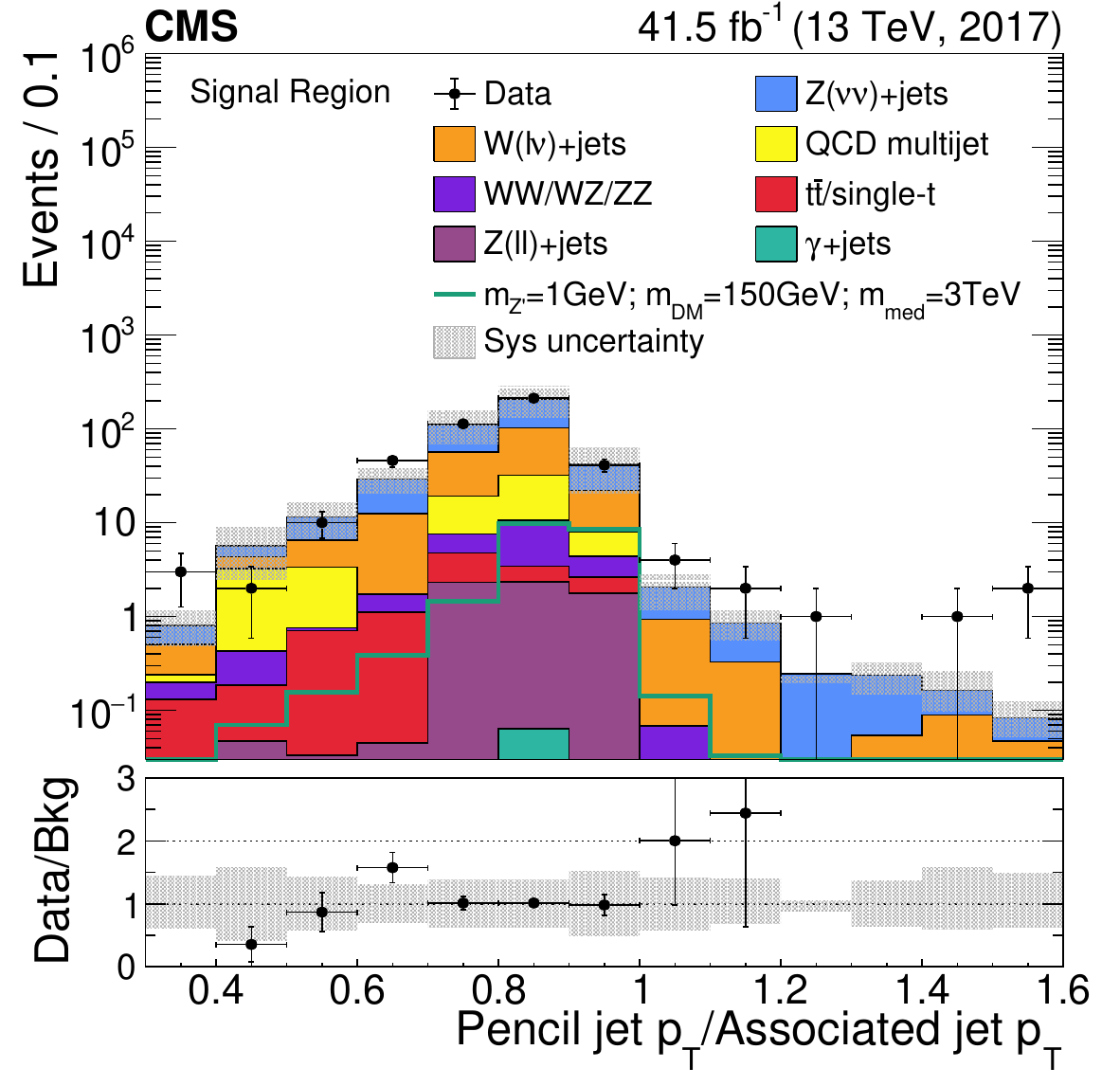}
      \caption{Comparison between observed data and simulated backgrounds in the signal region for the recoil (upper-left), the charged hadron energy fraction of the associated jet (upper-right), the neutral electromagnetic energy fraction of the associated jet (lower-left), and the ratio of pencil-jet \pt to the associated jet \pt (lower-right). Total systematic uncertainties are labeled as ``Sys uncertainity" in the legend box. The ratio of data to the pre-fit background prediction is shown in the bottom pads, with the grey band indicating the total systematic uncertainty.}
      \label{fig:sr_2017}
\end{figure*}

For each data-taking period (two for 2016, and one each for 2017 and 2018), four validation regions are defined to evaluate the agreement between data and simulation and to assess the performance of the machine learning (ML) model prior to unblinding the signal region. These regions correspond to single electron, single muon, double electron, and double muon event selections, constructed by replacing the lepton veto requirements used in the signal region with the corresponding lepton selection criteria. The single-lepton regions are enriched in $\PW\to\Pell\PGn$ events, while the dilepton regions are enriched in $\PZ\to\Pell\Pell$ events. Distributions of key variables in the single-electron validation region are shown in Fig.~\ref{fig:single_ele_cr_2017}, while those for the signal region are shown in Fig.~\ref{fig:sr_2017}, demonstrating good agreement between data and simulation within uncertainties. A similarly good level of agreement is observed in the other validation regions. To ensure sufficient statistics, the 2016 preVFP and postVFP data sets are combined for training the ML model and for producing the signal region distribution.

\begin{table*}[ht!]
   \topcaption{List of the input variables used in the ML model along with brief descriptions of their physical meaning.}
   \label{tab:ml_inputs}
   \centering
   \renewcommand{\arraystretch}{1.0}
   \begin{scotchx}{ll}
      Name of input variables & Description \\
      \hline
      \rule{0pt}{0.5cm}$\frac{\text{leading Tk \pt}}{\text{(pencil-jet) \pt}}$ & Ratio of \pt of leading track of pencil-jet and the pencil-jet \pt \\[\cmsTabSkip]
      $\Delta{R}$(pencil-jet, leading Tk) & $\Delta{R}$ separation between leading track of pencil-jet and pencil-jet \\
      $\text{pencil-jet}_\text{mass}$& Mass of the pencil-jet as calculated from the constituents \\[\cmsTabSkip]
      $\frac{\text{leading Tk}\ \pt}{\text{associated\ jet \pt}}$ & Ratio of \pt of leading track of pencil-jet and the associated jet \pt (defined in section ~\ref{sec:detector}) \\
      $\frac{\text{(pencil-jet) \pt}}{\text{associated jet \pt}}$ & Ratio of \pt of pencil-jet and the associated jet \pt \\
      $\Delta$R(pencil-jet, associated jet)& $\Delta$R separation between pencil-jet and associated jet \\[\cmsTabSkip]
      associated $\text{jet}_\text{chHEF}$ & Fraction of energy carried by charged hadrons in the associated jet  \\
      associated $\text{jet}_\text{neEmEF}$ & Fraction of energy carried by photons in the associated jet \\
      associated $\text{jet}_\text{neHEF}$ & Fraction of energy carried by neutral hadrons in the associated jet \\
      associated $\text{jet}_\text{mass}$& Mass of the associated jet as calculated from the constituents \\
      recoil& PF \ptmiss \\
      $\Delta \phi$(pencil-jet, recoil) & $\Delta \phi$ separation between the pencil-jet and recoil \\
      $\text{(pencil-jet)}_\text{decaymode}$ & Encodes the number of prongs in the pencil-jet \\ 
      \end{scotchx}
\end{table*}

\section{Signal extraction strategy}
\label{sec:ml_model}

After the event selection is applied as described in Section~\ref{sec:selection}, the remaining background is dominated by processes with kinematic properties similar to those of the signal, particularly $\Zvvjets$ and \Wlvjets production. The discrimination between signal and background relies on subtle differences in jet substructure, event kinematics, and correlations among multiple observables, that traditional cut-based approaches do not fully exploit. To enhance the sensitivity of the analysis, an ML approach is therefore employed to combine these observables into a single powerful discriminant.

The ML model is trained using a set of input variables designed to capture the distinctive features of the \PZpr boson. These variables include kinematic properties of the pencil-jet and its associated jet, as well as global event quantities such as the missing transverse momentum (recoil). Particular emphasis is placed on variables sensitive to the low-multiplicity and collimated nature of the pencil-jet, including track-based observables and jet substructure quantities. 
The input variables are chosen based on both physical intuition and empirical studies, prioritizing features that provide strong discrimination power, while remaining robust against pileup effects and potential mismodelling in simulation. A complete list of the input features is provided in Table~\ref{tab:ml_inputs}.

A two-stage ML strategy is adopted following a comparative evaluation of several architectures, including neural networks (NNs) and boosted decision trees (BDTs). In the first stage, a feed-forward NN is implemented using TensorFlow~\cite{Abadi:2016kic}. The network consists of three hidden layers with 48, 36, and 3 nodes, respectively, and employs parametric ReLU activation functions along with dropout regularization, to mitigate overfitting. The training is performed using a binary cross-entropy loss function, optimized with the adaptive moments algorithm. The network contains a total of 2635 trainable parameters.
In the second stage, the output of the NN is combined with the original input variables and used as input to a BDT implemented with XGBoost~\cite{xgboost-paper}. The BDT is configured with 250 estimators, each trained on a random subset of 10 out of the 14 available features to improve robustness. A maximum tree depth of 10 is used, and the model is trained with a logistic regression objective function. The final BDT output, hereafter referred to as the ML output, serves as the primary discriminant for signal extraction.

To ensure uniform performance across different signal mass hypotheses, all signal mass points are combined into a single signal class, while all background processes are grouped into a single background class and weighted according to their production cross section times acceptance.
A subset of the simulated samples (approximately 12--13\% of signal and background events) is used for training. 

The hyperparameters of both models are optimized using a randomized search over a predefined grid. Among the input variables, the recoil, the fraction of energy carried by photons in the associated jet; and the ratio of the transverse momentum of the pencil-jet to that of the associated jet are found to have the largest impact on the model performance. The variable importance is evaluated using a permutation-based approach, in which each variable is randomly shuffled across 100 replicas of the dataset, and the resulting degradation in model performance is used to rank the features.

\begin{figure*}[htb!]
   \centering
      \includegraphics[width=\textwidth]{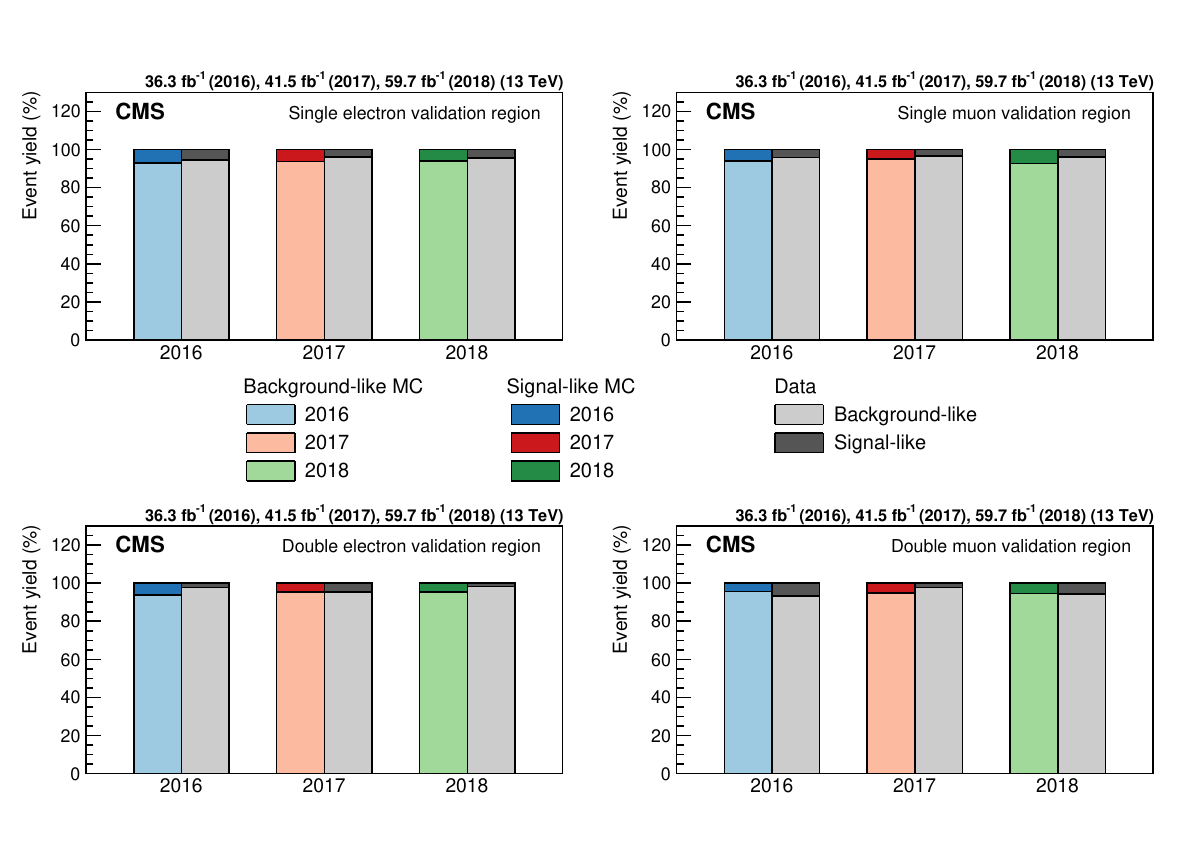}
      \caption{Percentage of events in the validation regions: single electron (upper-left), single muon (upper-right), double electron (lower-left), and double muon (lower-right), with low and high ML scores. The lighter shading represents the fraction of events with low scores (as expected for these background-enriched regions), while the darker shading corresponds to the fraction with high scores ($>0.71$), indicative of signal-like behavior. Colored markers denote simulated samples for 2016 (blue), 2017 (red), and 2018 (green), while data are shown in grey.
        }
      \label{fig:ml_performance_CR}  
\end{figure*}

\begin{figure*}[htbp]
   \centering
      \includegraphics[width=0.49\textwidth]{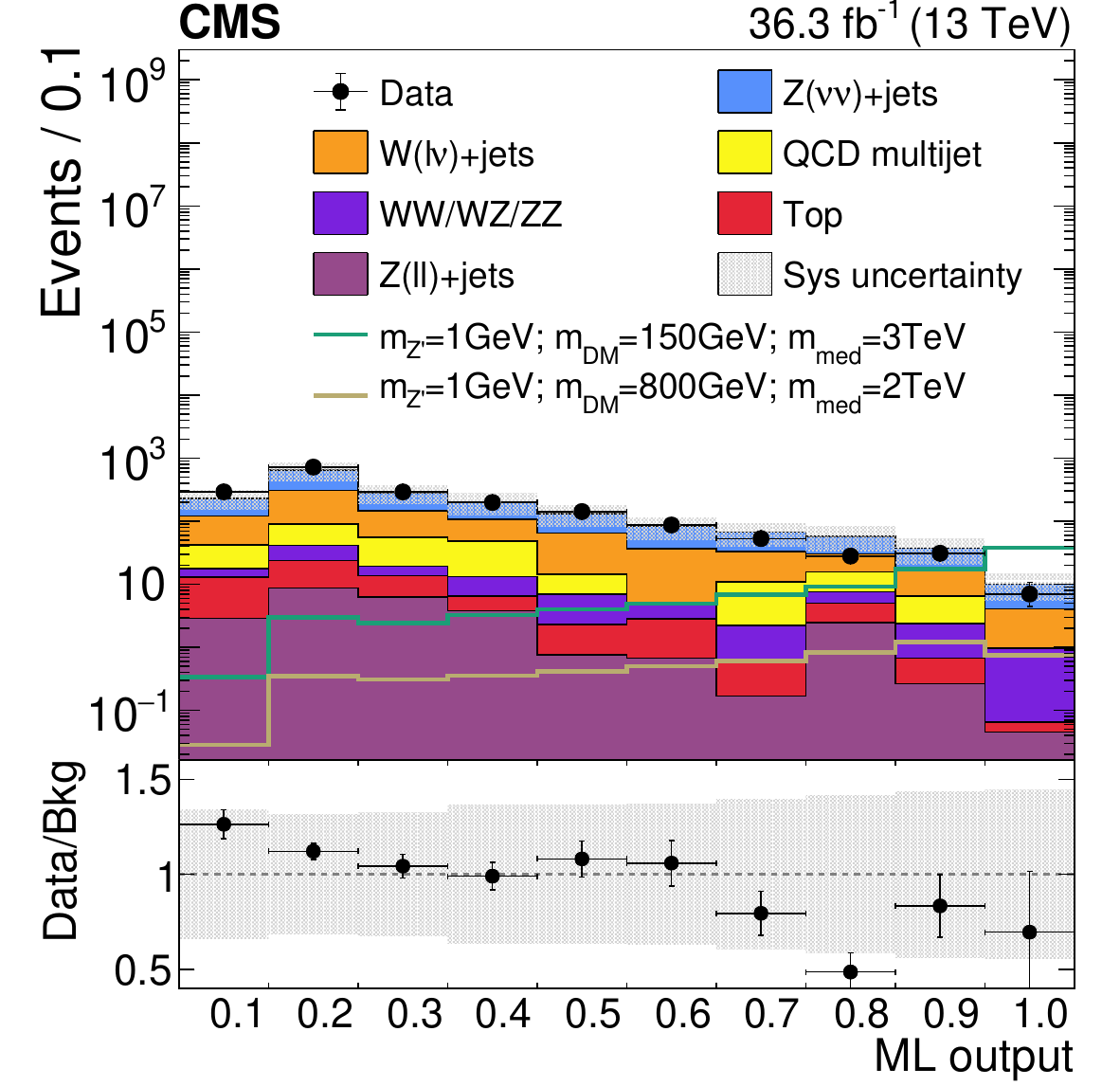}
      \includegraphics[width=0.49\textwidth]{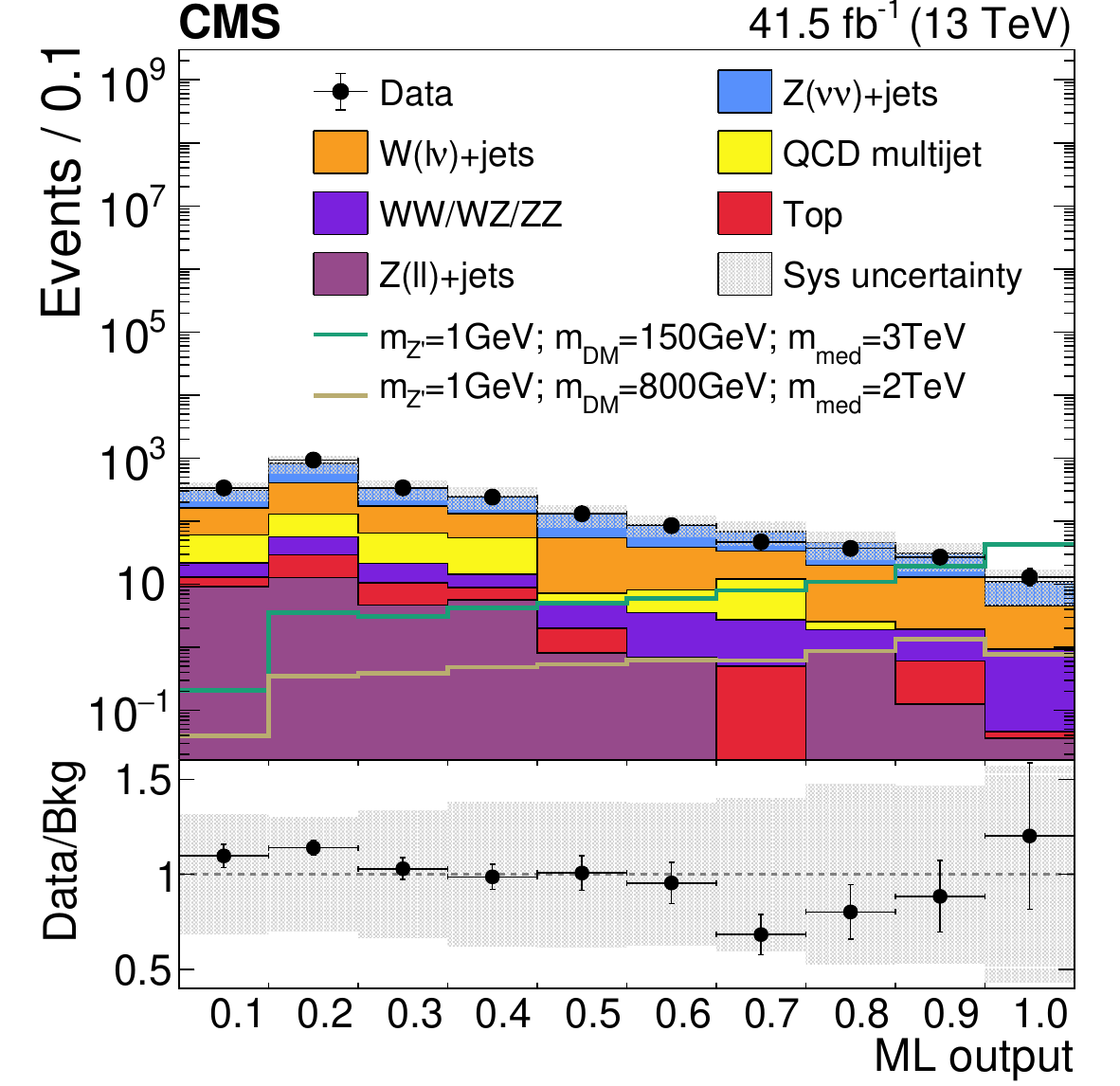}
      \includegraphics[width=0.49\textwidth]{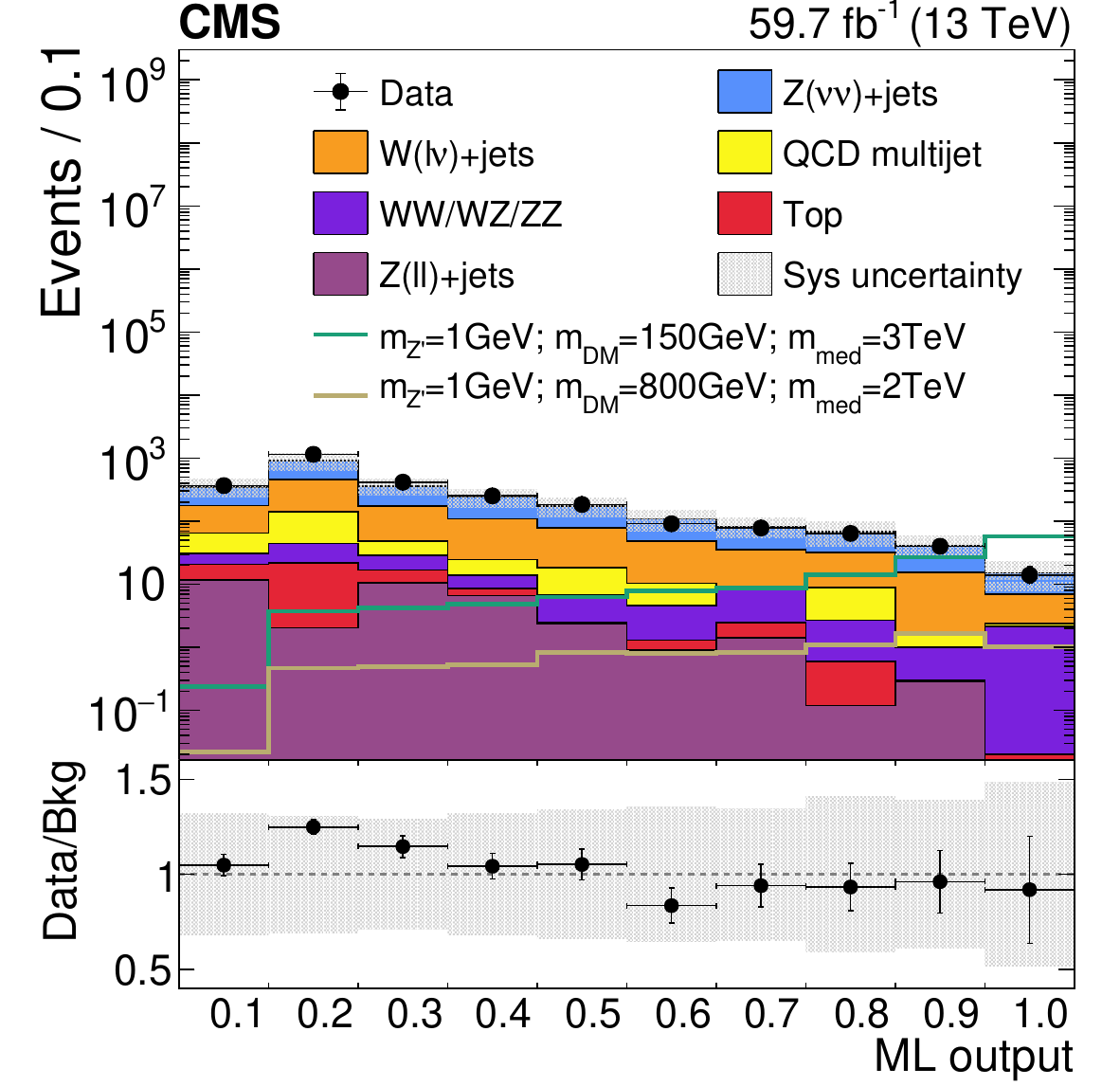} \\
      \caption{Pre-fit distributions of the ML output after full unblinding for the three data-taking periods: 2016 (upper-left), 2017 (upper-right), and 2018 (lower), comparing data with the total expected background. Distributions for two representative signal samples are overlaid, normalized to their expected theoretical cross sections. In the legend, ``Sys uncertainty" denotes total systematic uncertainties. The bottom pad in each plot shows the ratio of data to the predicted background, with the gray band indicating the pre-fit uncertainty.}
      \label{fig:prefit}
\end{figure*}

To improve robustness against measurement uncertainties, a data augmentation strategy is employed. ``Toy" events are generated by independently smearing each input variable according to its estimated pre-fit uncertainty, with the exception of the pencil-jet decay mode, which is kept fixed. The model is trained on a small fraction of the 2017 simulated samples (less than 15\%) and subsequently evaluated on independent simulated and data samples from the 2016, 2017, and 2018 data-taking periods. The consistent performance observed across all eras, both in validation regions (Fig.~\ref{fig:ml_performance_CR}) and signal regions (Fig.~\ref{fig:prefit}), demonstrates that the model successfully captures the underlying signal characteristics while remaining robust against variations in detector conditions and simulation mismodelling.

The performance of the ML model is validated in multiple validation regions enriched in leptonic processes, including single-electron, single-muon, and dilepton event samples. These regions are designed to be orthogonal to the signal region and provide a stringent test of the modelling of both the input variables and the ML response. A wide range of input variable distributions are compared between data and simulation, with good agreement observed within uncertainties. Representative distributions, such as recoil, charged hadron energy fraction, neutral electromagnetic energy fraction of the associated jet and the ratio of pencil-jet \pt to the associated jet \pt in the single electron validation region are shown in Fig.~\ref{fig:single_ele_cr_2017}.  

The ML output distributions are also examined in these regions, showing consistent agreement between data and simulation as summarized in Fig.~\ref{fig:ml_performance_CR}. Figure~\ref{fig:ml_performance_CR} shows the percentage of events in the validation region with low ($<0.71$) versus high ($>0.71$) ML scores for both data and simulation. The fraction of events with high scores, indicative of signal-like behavior, is found to differ by less than 5\%, across all validation samples and data-taking periods, which is well within the pre-fit uncertainties. This level of agreement provides confidence that the ML model is accurately capturing the relevant physics features and that the simulation is sufficiently modelling the input variables and their correlations. The consistent performance across different eras and validation regions supports the robustness of the model and its suitability for application to the signal region for the final statistical analysis.

Following this validation, the trained ML model is applied to the full dataset. Each event is assigned a continuous discriminant score in the range [0, 1], with larger values indicating a higher likelihood of being signal-like. The resulting ML output distributions in the signal region, shown in Fig.~\ref{fig:prefit}, are used as inputs to the final statistical analysis. These distributions exhibit good agreement between data and simulation while providing enhanced discrimination between signal and background, forming the basis of the likelihood fit described in Section~\ref{sec:combine}.

\section{Sources of uncertainty}
\label{sec:uncertainty}
Both theoretical and experimental sources of systematic uncertainty are considered for the signal and background processes, as described in this section. The impact of each uncertainty on the shape of the input features is evaluated and subsequently propagated through the ML model to determine its effect on the final classifier output. A summary of all considered systematic uncertainties is provided in Table~\ref{tab:all_sys}, where the quoted values represent the corresponding variations in the event yield expressed as percentages.

\subsection{Experimental sources}
\label{ssec:expt_unc}
Dominant experimental uncertainties arise from the reconstruction of the pencil-jet object. As described in Section~\ref{sec:selection}, these uncertainties account for potential mismodelling in both the selection efficiency and misidentification rates of the pencil-jet in data relative to simulation. Their impact on the event yield typically ranges 12--29\%, with the largest contributions originating from limited statistics of genuine \PGt leptons reconstructed in the two-prong decay mode. Corresponding uncertainties are generally smaller for the one- and three-prong categories.
Additional experimental uncertainties include those associated with the JES and JER, each contributing a the level of approximately 5\% to the event yield. The trigger efficiency is evaluated using a data sample enriched in single-muon events, leading to an associated uncertainty in the scale factor of about 10\%. Smaller contributions from uncertainties in the integrated luminosity, prefiring effects~\cite{Hayrapetyan_2024}, and the efficiency of b-tagged jet identification are also taken into account.

\subsection{Theoretical sources}
Theoretical uncertainties affecting the \zjets and \wjets processes arise primarily from missing higher-order corrections in perturbative QCD and EW calculations. These uncertainties are summarized in Table~\ref{tab:all_sys} and described below.

\begin{table*}[ht!]
\centering
\topcaption{Major sources of systematic uncertainty. All listed uncertainties affect the shapes of the input variables. The ``Value" column indicates the impact of each uncertainty, expressed as a percentage of the event yield. Most uncertainties are correlated across processes and data-taking eras, except for the pencil-jet scale factors and misidentification rates, which are treated as uncorrelated across decay modes and eras. In addition, misidentification uncertainties are taken to be correlated among non-QCD processes. The labels ``QCD" and ``EW" denote uncertainties associated with higher-order QCD and electroweak effects, respectively. They affect the \zjets and \wjets processes.}
\begin{scotch}{ll}
      Source of systematic uncertainty & Value (\%)   \\
      \hline
      Fact. \& renorm. scales (QCD) & 8--26 \\
      $\pt$ shape dependence (QCD)   & 1--25  \\
      Process dependence (QCD)   & ${\approx}2.5$  \\
      Effects of unknown Sudakov logs (EW)   & ${\approx}5$ \\
      Missing NNLO effects (EW)   & ${\approx}2.5$  \\
      Effects of NLL Sudakov approximations (EW)   & ${\approx}2.2$  \\
      Unfactorized mixed QCD-EW corrections   & ${\approx}2.7$ \\
      PDF modelling   & ${\approx}5$ \\
      Pencil-jet SF & 12--29   \\
      Pencil-jet misidentification    & 20--25   \\
      JES and JER   & ${\approx}5$ \\
\end{scotch}
\label{tab:all_sys}
\end{table*}

\begin{figure*}[htbp]
   \centering
      \includegraphics[width=0.49\textwidth]{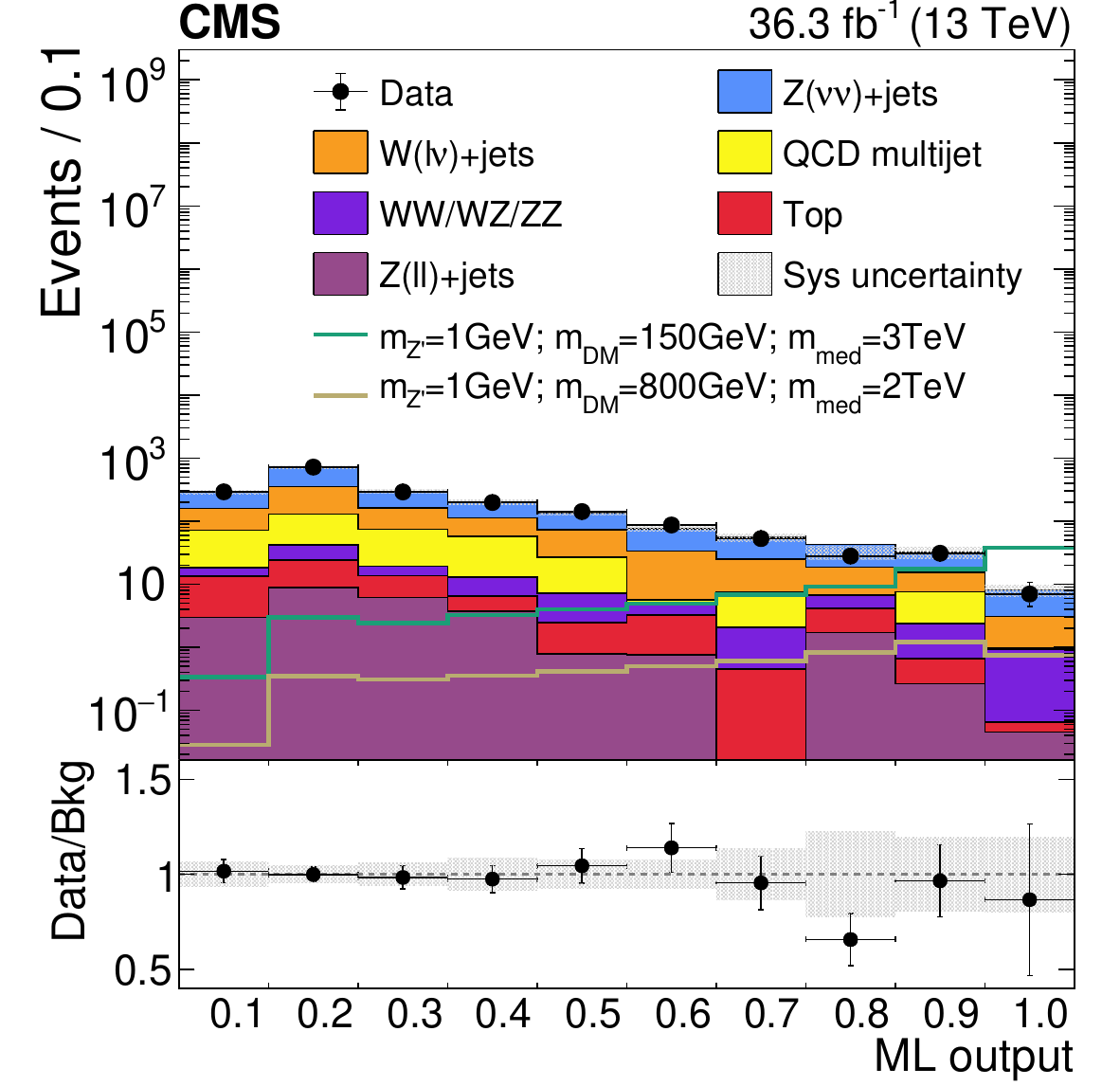}
      \includegraphics[width=0.49\textwidth]{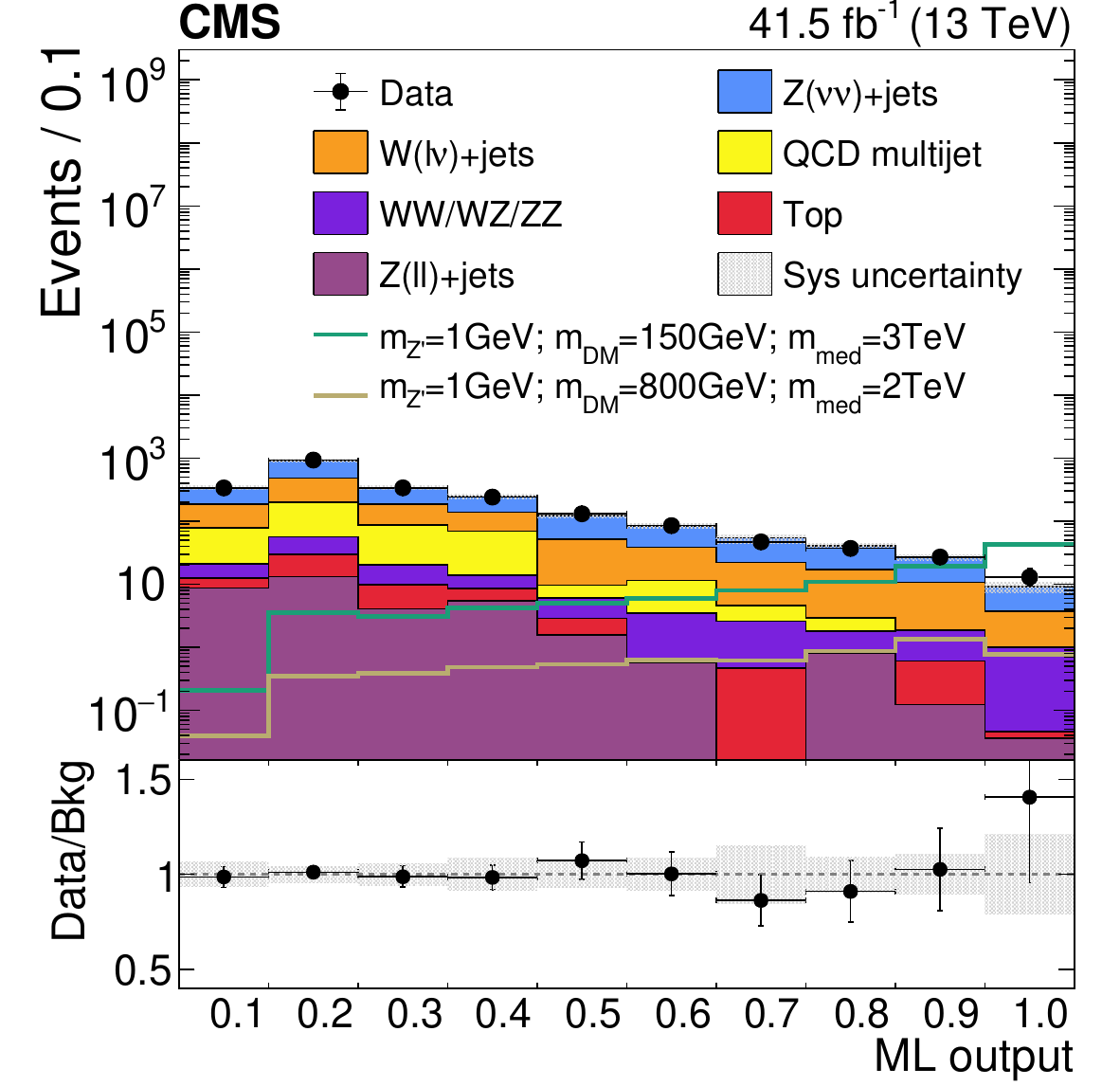}
      \includegraphics[width=0.49\textwidth]{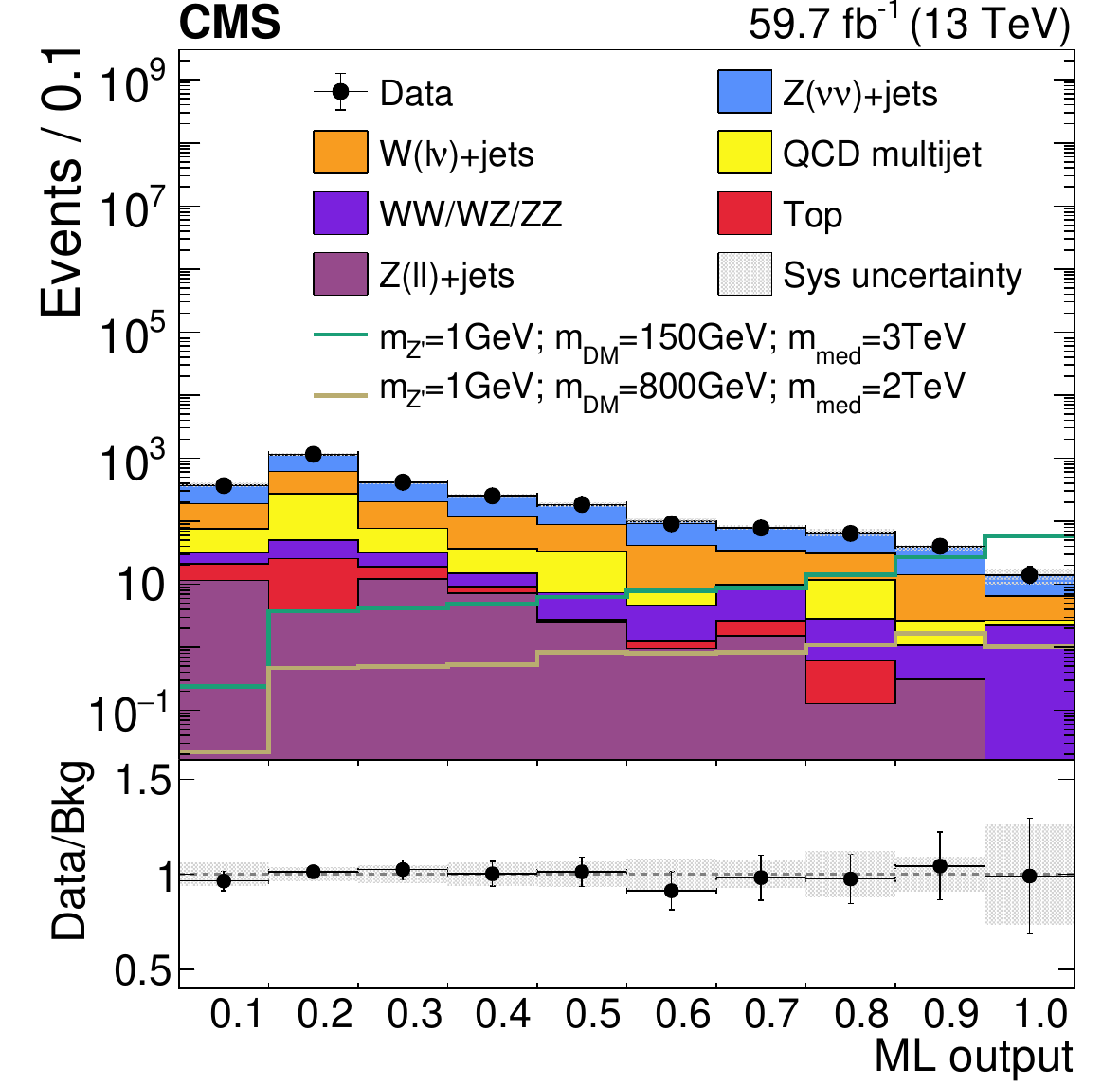} 
      \caption{Post-fit distributions of the ML output for the three data-taking periods: 2016 (upper-left), 2017 (upper-right), and 2018 (lower), comparing data with the total expected background. Distributions for two representative signal samples are overlaid, normalized to their expected theoretical cross sections. In the legend, ``Sys uncertainty" denotes total systematic uncertainties. The bottom pad in each plot shows the ratio of data to the predicted background, with the gray band indicating the post-fit uncertainty.}
      \label{fig:postfit}
\end{figure*}

Three independent sources are considered to account for QCD higher-order effects. The first corresponds to variations of the renormalization and factorization scales, evaluated by changing each scale by factors of two and one half relative to its nominal values. These variations mainly affect the overall normalization of the boson \pt spectrum but do not fully capture potential shape distortions, particularly in the high-\pt region. To address this limitation, a second, more conservative shape uncertainty is introduced based on alternative boson \pt spectra, providing additional sensitivity to \pt-dependent effects. A third uncertainty accounts for differences in higher-order QCD corrections between \wjets and \zjets production, derived from the relative variation of their corresponding correction factors.

At high energies, EW corrections are dominated by large Sudakov logarithms of the form $\log^2(s/M_{\PW,\PZ}^2)$, arising from soft and collinear gauge boson emissions. These effects can lead to significant negative corrections to cross sections and must be accounted for. To this end, resummation-based approximations are employed, and associated uncertainties are included following the prescription in Ref.~\cite{DMTheory}. Three independent EW uncertainty sources are considered: missing higher-order Sudakov logarithms beyond next-to-next-to-leading order (NNLO), incomplete NNLO contributions not captured in approximate calculations, and differences between the next-to-leading logarithmic (NLL) Sudakov treatment at two loops and the exponentiation of full next-to-leading order (NLO) EW corrections. In addition, mixed QCD--EW corrections are incorporated as described in Ref.~\cite{DMTheory}.

\begin{figure}[ht!]
   \centering
      \includegraphics[width=0.49\textwidth]{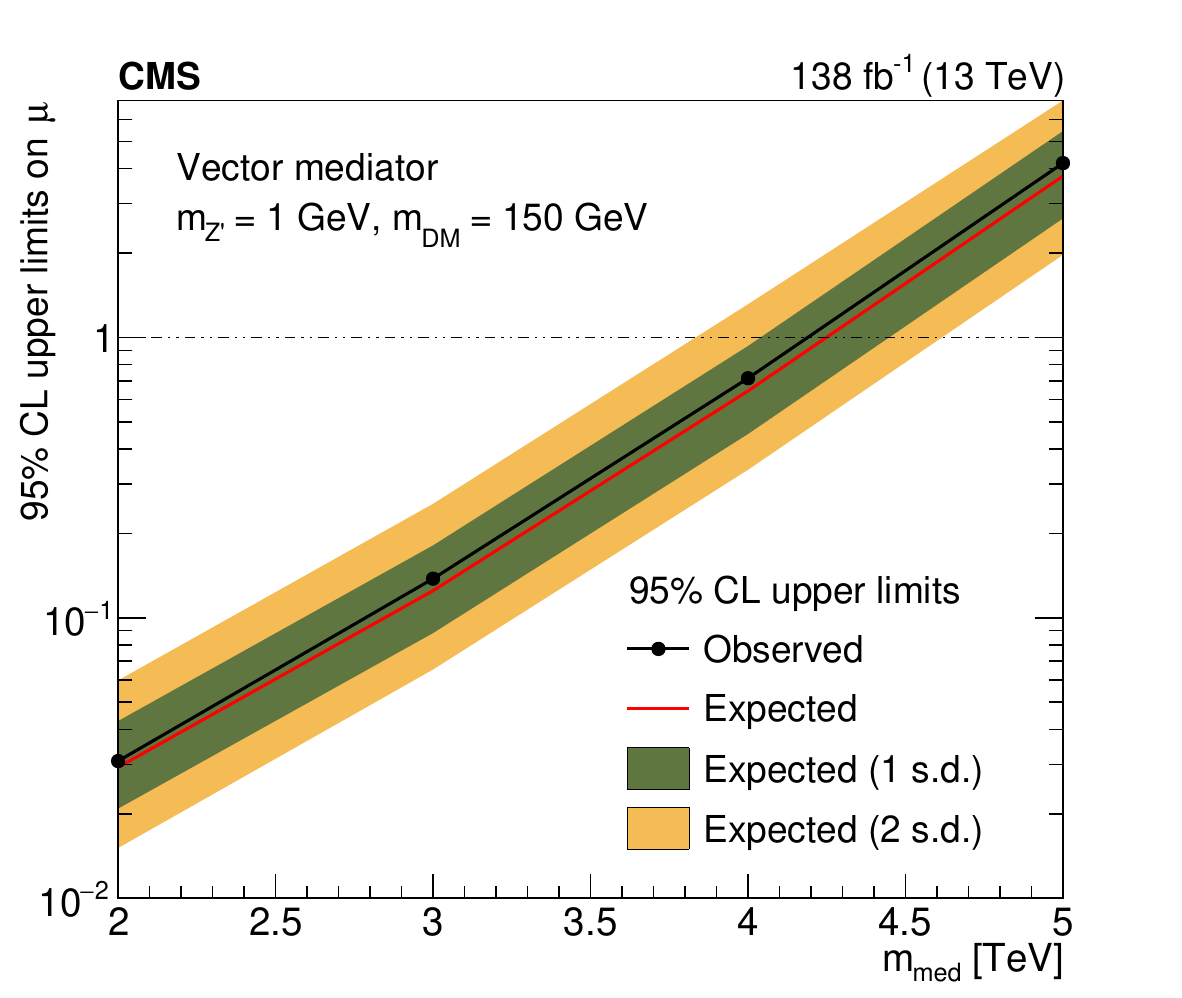}
      \includegraphics[width=0.49\textwidth]{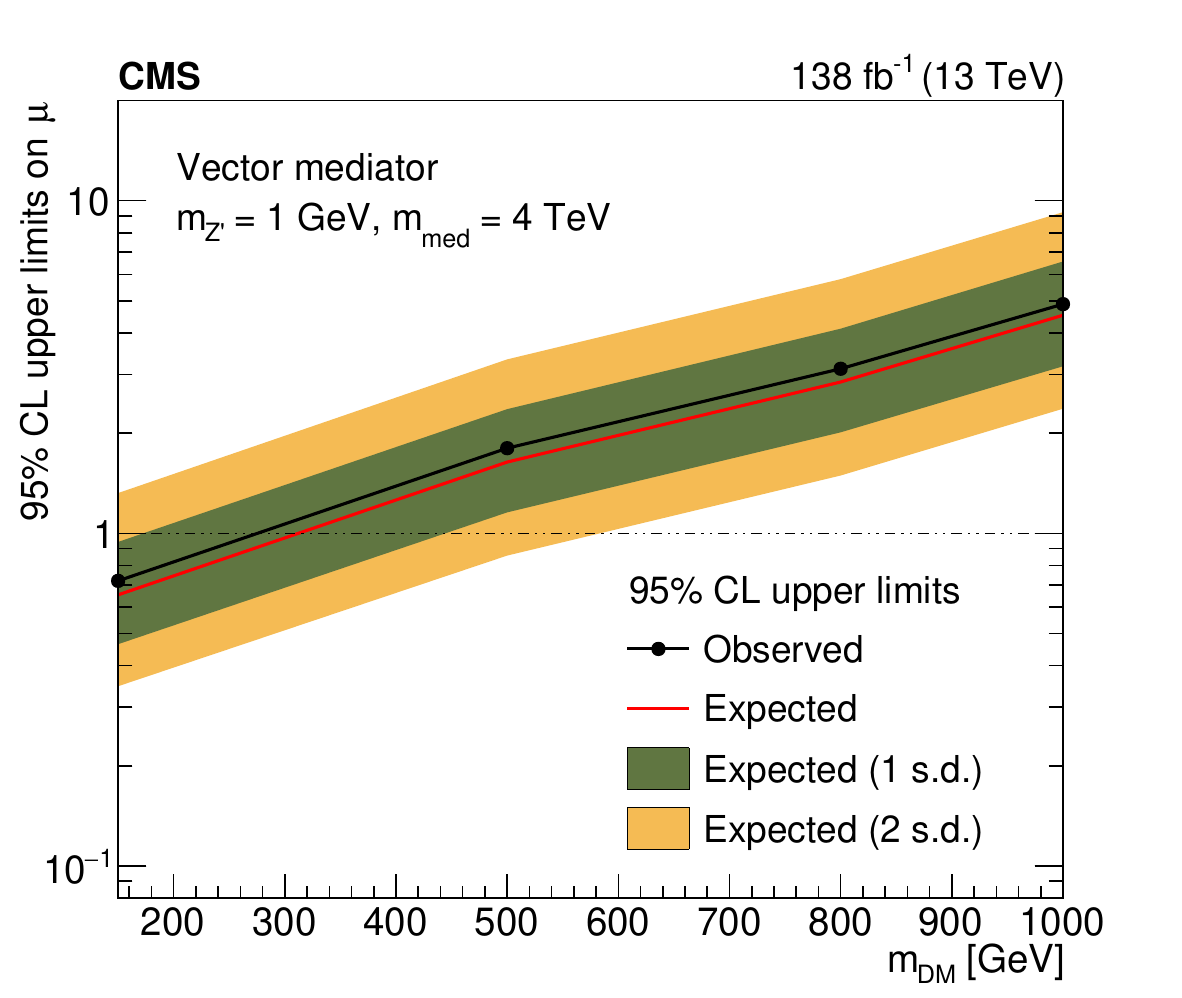}
      \caption{The 95\% \CL upper limits on $\mu = {\sigma}/{\sigma_{\text{theory}}}$ for the vector mediator with $\mZpr = 1\GeV$. The \cmsLeft panel shows the limits as a function of the mediator mass, $\mmed$ for a fixed dark matter mass of $\mdm=150\GeV$, while the \cmsRight panel shows the limits as a function of the dark matter mass, $\mdm$ for a fixed mediator mass of $\mmed = 4\TeV$. A linear interpolation is applied between the available mass points. The inner (green) and outer (yellow) bands correspond to the 1 and 2 standard deviations (s.d.), respectively, of the expected limits under the background-only hypothesis.
      }
      \label{fig:fitbased_limits_vector}
\end{figure}

Uncertainties related to the choice of parton distribution functions (PDFs) are evaluated following the recommendations of Ref.~\cite{PDF4LHCWorkingGroup:2022cjn}. For the signal processes, an additional normalization uncertainty of 10\% is assigned to account for variations in the QCD renormalization and factorization scales. Contributions from minor background processes, including QCD multijet, diboson, and top quark production, are also assigned corresponding yield uncertainties.

Finally, uncertainties arising from the finite size of simulated samples used in the fit are incorporated using the Barlow--Beeston--lite method~\cite{BARLOW1987202}.

\section{Results}
\label{sec:combine}
A binned maximum-likelihood signal-plus-background fit is performed on the ML classifier output using the full dataset collected across the three data-taking years, employing the CMS statistical analysis framework, \textsc{Combine}~\cite{CMS:published_cms_combine}. The fit is carried out exclusively in the signal region, while validation regions are used independently to validate the modelling of input variables and the ML response, as described in previous sections. Systematic uncertainties are incorporated as nuisance parameters, with their correlations properly taken into account.  No statistically significant excess above the SM expectation is observed.

\begin{figure}[ht!]
   \centering
      \includegraphics[width=0.49\textwidth]{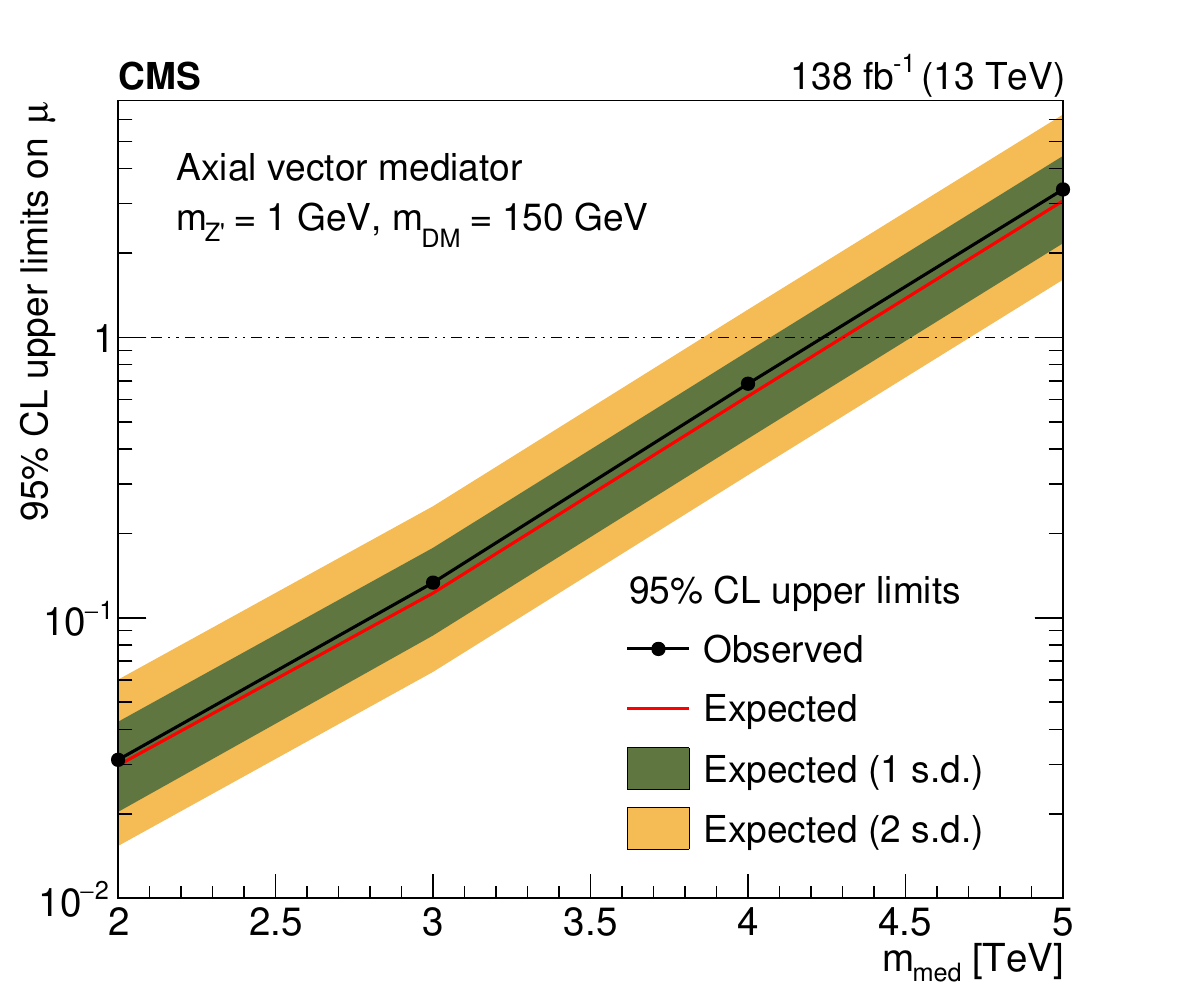}
      \includegraphics[width=0.49\textwidth]{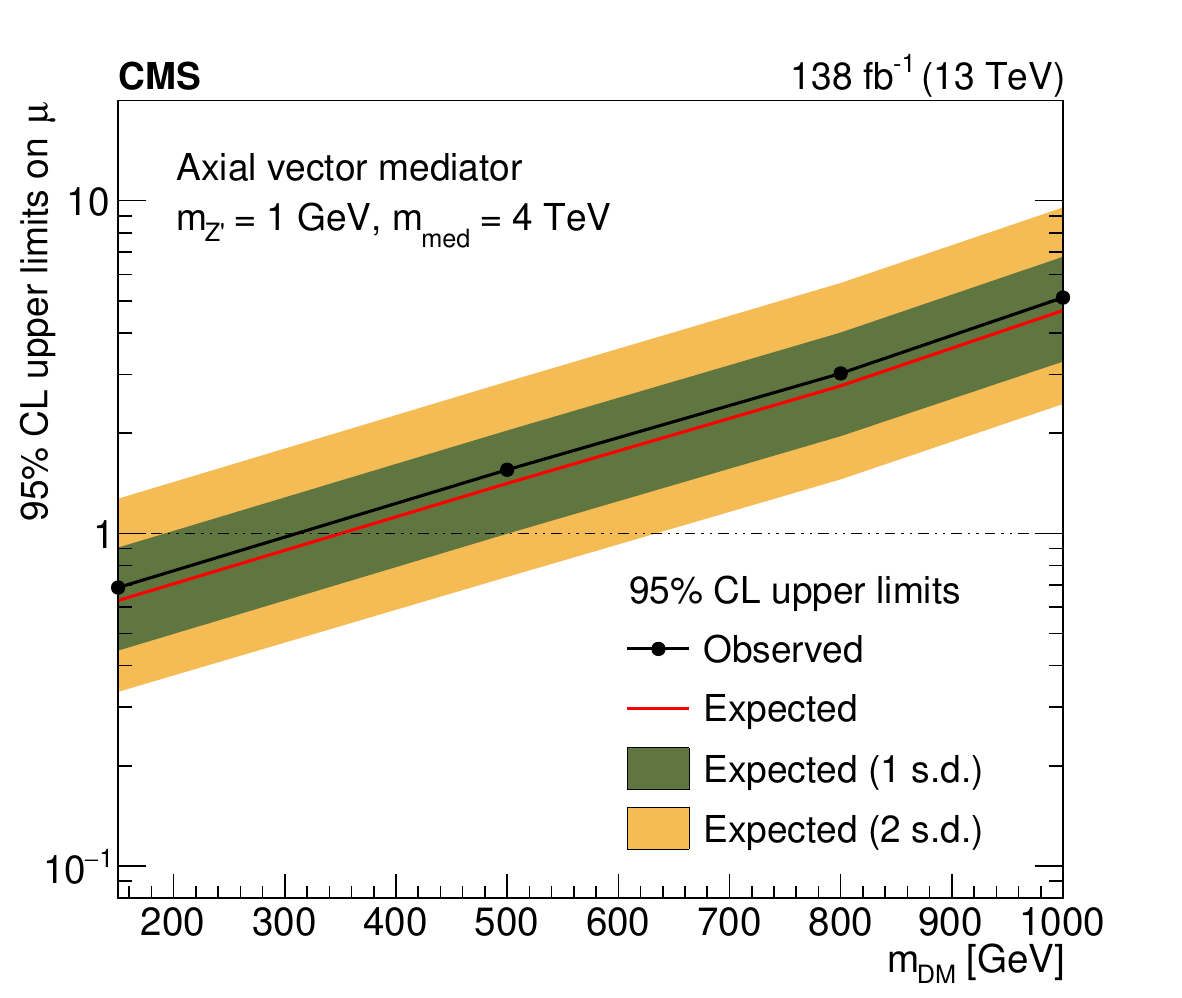}
      \caption{The 95\% \CL upper limits on $\mu = {\sigma}/{\sigma_{\text{theory}}}$ for the axial vector mediator with $\mZpr = 1\GeV$. The \cmsLeft panel shows the limits as a function of the mediator mass, $\mmed$ for a fixed dark matter mass of $\mdm=150\GeV$, while the \cmsRight panel shows the limits as a function of the dark matter mass, $\mdm$ for a fixed mediator mass of $\mmed = 4\TeV$. A linear interpolation is applied between the available mass points. The inner (green) and outer (yellow) bands correspond to the 1 and 2 standard deviations (s.d.), respectively, of the expected limits under the background-only hypothesis.}
      \label{fig:fitbased_limits_axial}
\end{figure}

Post-fit distributions of the ML output are shown in Fig.~\ref{fig:postfit}, where good agreement is observed between data and the SM background predictions within the estimated uncertainties. The overall sensitivity of the analysis is primarily limited by the statistical precision of the data, with systematic uncertainties contributing at a subleading level.

The results are interpreted in terms of constraints on dark matter (DM) production in simplified models with vector and axial-vector mediators. Upper limits are set on the signal strength modifier, $\mu = {\sigma}/{\sigma_{\text{theory}}}$
where $\sigma_{\text{theory}}$ is the predicted production cross section for a given signal hypothesis. The theoretical cross sections are calculated using \MGvATNLO (version 2.6.5), based on the model assumptions and parameter choices described in Section~\ref{sec:intro}. The limits are derived at 95\% confidence level (\CL) using the modified frequentist \CLs method~\cite{CLS1,CLS2}, with the test statistic evaluated using asymptotic approximations~\cite{CLS3}.

\begin{figure}[ht!]
   \centering
      \includegraphics[width=0.49\textwidth]{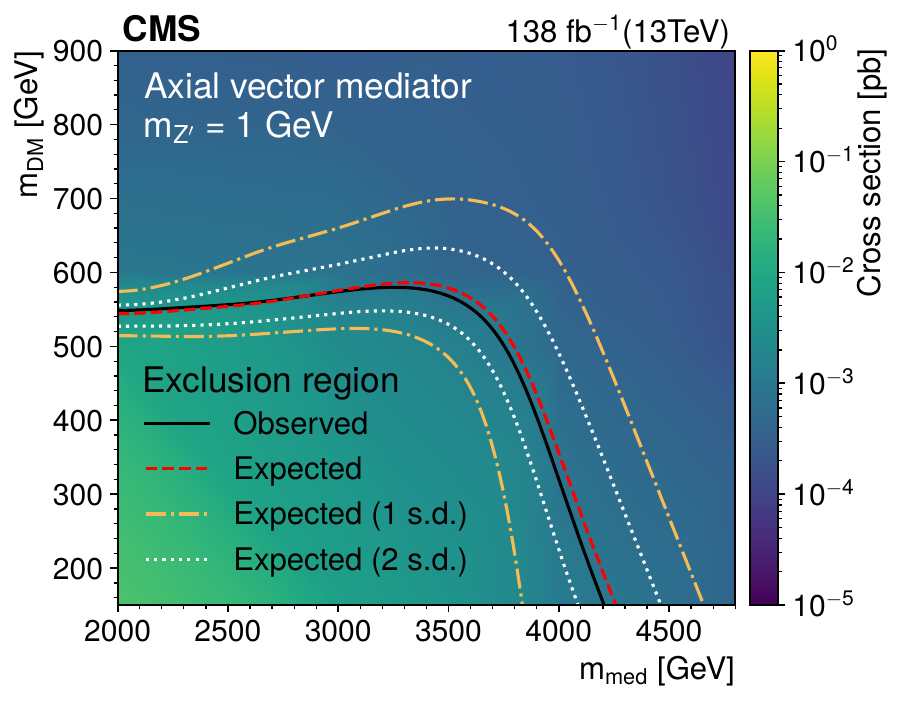}
      \includegraphics[width=0.49\textwidth]{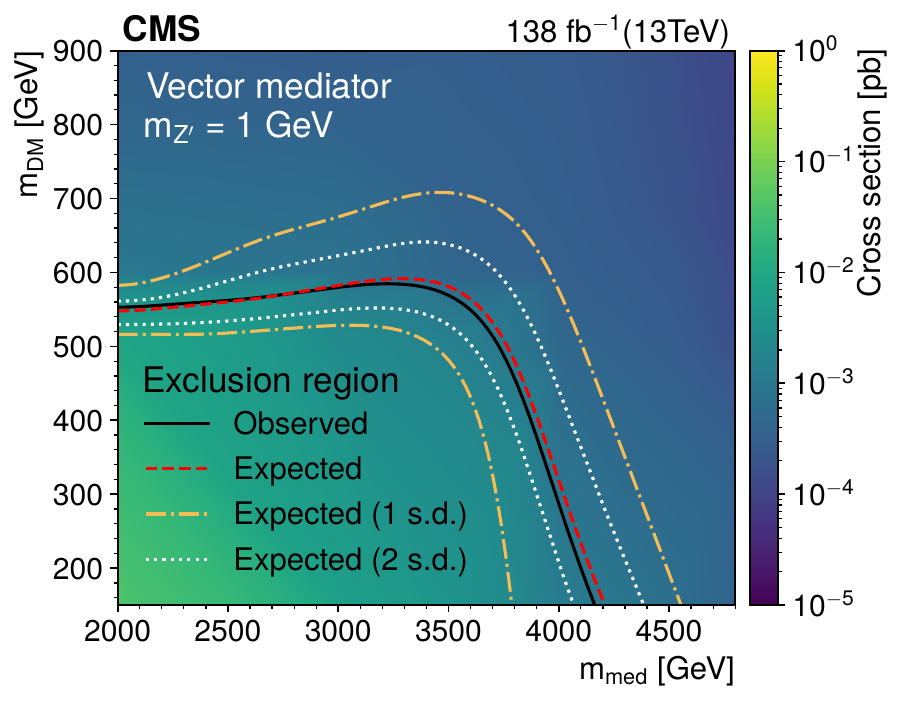}
      \caption{Expected and observed exclusion regions in the $\mdm-\mmed$ plane for the axial vector mediator (\cmsLeft) and vector mediator (\cmsRight), assuming $\mZpr=1\GeV$. The parameter space in the lower-left region of each panels is excluded at the 95\% \CL level. The color scale indicates the theoretical cross section of the signal process.}
      \label{fig:fitbased_2d_limits}
\end{figure}

The results obtained with the full dataset are presented in Figs.~\ref{fig:fitbased_limits_vector} and~\ref{fig:fitbased_limits_axial}, showing the observed and expected limits as functions of the mediator mass for fixed DM mass hypotheses, and vice versa for both vector and axial-vector mediator scenarios. The corresponding exclusion contours in the $\mdm-\mmed$ plane are shown in Fig.~\ref{fig:fitbased_2d_limits}.

For DM masses of 150\GeV, mediator masses up to about 4.25\TeV are excluded at 95\% \CL in both the vector and axial-vector cases. For higher DM masses around 550\GeV, the exclusion reach extends up to mediator masses of approximately 3.5\TeV. The reduced sensitivity at higher masses is primarily because of the rapidly decreasing signal production cross section, which results in fewer expected signal events.

No evidence for DM production is observed, and the results place stringent constraints on simplified DM models with spin-1 mediators. The analysis demonstrates strong sensitivity to high-mass mediator scenarios and underscores the effectiveness of the ML-based approach in enhancing signal discrimination. While no evidence for DM production is observed, the derived limits provide important guidance for future searches, particularly in regions of phase space where improvements in luminosity and analysis techniques could further extend the sensitivity.

\section{Summary}
\label{sec:summary}
A novel search for dark matter (DM) particles is presented using 138\fbinv of proton-proton collision data recorded by the CMS detector at a center-of-mass energy of 13\TeV. The analysis targets a distinctive final state characterized by a large transverse momentum imbalance and an energetic low-multiplicity jet. This topology provides a distinctive and potentially more sensitive probe of dark-sector dynamics than traditional mono-jet searches. To enhance sensitivity to the signal, supervised machine learning techniques are employed. No significant deviation from the standard model predictions is observed. Upper limits at 95\% confidence level are set on the DM production cross section within the framework of simplified models with vector and axial-vector mediators. Mediator masses up to approximately 4.25\TeV are excluded for a DM mass of 150\GeV, while for a DM mass of 550\GeV, mediator masses up to approximately 3.5\TeV are excluded in both scenarios. These results provide the first constraints on this novel final state and significantly extend the sensitivity of DM searches at the LHC. 

\begin{acknowledgments}
We congratulate our colleagues in the CERN accelerator departments for the excellent performance of the LHC and thank the technical and administrative staffs at CERN and at other CMS institutes for their contributions to the success of the CMS effort. In addition, we gratefully acknowledge the computing centers and personnel of the Worldwide LHC Computing Grid and other centers for delivering so effectively the computing infrastructure essential to our analyses. Finally, we acknowledge the enduring support for the construction and operation of the LHC, the CMS detector, and the supporting computing infrastructure provided by the following funding agencies: SC (Armenia), BMFWF and FWF (Austria); FNRS and FWO (Belgium); CNPq, CAPES, FAPERJ, FAPERGS, and FAPESP (Brazil); MES and BNSF (Bulgaria); CERN; CAS, MoST, and NSFC (China); MINCIENCIAS (Colombia); MSES and CSF (Croatia); RIF (Cyprus); SENESCYT (Ecuador); ERC PRG and PSG, TARISTU24-TK10 and MoER TK202 (Estonia); Academy of Finland, MEC, and HIP (Finland); CEA and CNRS/IN2P3 (France); SRNSF (Georgia); BMFTR, DFG, and HGF (Germany); GSRI (Greece); MATE and NKFIH (Hungary); DAE and DST (India); IPM (Iran); SFI (Ireland); INFN (Italy); MSIT and NRF (Republic of Korea); MES (Latvia); LMTLT (Lithuania); MOE and UM (Malaysia); BUAP, CINVESTAV, CONACYT, LNS, SEP, and UASLP-FAI (Mexico); MOS (Montenegro); MBIE (New Zealand); PAEC (Pakistan); MSHE, NSC, and NAWA (Poland); FCT (Portugal);  MESTD (Serbia); MICIU/AEI and PCTI (Spain); MOSTR (Sri Lanka); Swiss Funding Agencies (Switzerland); MST (Taipei); MHESI (Thailand); TUBITAK and TENMAK (T\"{u}rkiye); NASU (Ukraine); STFC (United Kingdom); DOE and NSF (USA).

\begin{sloppypar}
\setlength\emergencystretch{\hsize}
\hyphenation{Rachada-pisek} Individuals have received support from the Marie-Curie program and the European Research Council and Horizon 2020 Grant, contract Nos.\ 675440, 724704, 752730, 758316, 765710, 824093, 101115353, 101002207, 101001205, and COST Action CA16108 (European Union); the Leventis Foundation; the Alfred P.\ Sloan Foundation; the Alexander von Humboldt Foundation; the Science Committee, project no. 22rl-037 (Armenia); the Fonds pour la Formation \`a la Recherche dans l'Industrie et dans l'Agriculture (FRIA) and Fonds voor Wetenschappelijk Onderzoek contract No. 1228724N (Belgium); the Beijing Municipal Science \& Technology Commission, No. Z191100007219010, the Fundamental Research Funds for the Central Universities, the Ministry of Science and Technology of China under Grant No. 2023YFA1605804, the Natural Science Foundation of China under Grant No. 12535004, and USTC Research Funds of the Double First-Class Initiative No.\ YD2030002017 (China); the Ministry of Education, Youth and Sports (MEYS) of the Czech Republic; the Shota Rustaveli National Science Foundation (Georgia); the Deutsche Forschungsgemeinschaft (DFG), among others, under Germany's Excellence Strategy -- EXC 2121 ``Quantum Universe" -- 390833306, and under project number 400140256 - GRK2497; the Hellenic Foundation for Research and Innovation (HFRI), Project Number 2288 (Greece); the Hungarian Academy of Sciences, the New National Excellence Program - \'UNKP, the NKFIH research grants K 131991, K 138136, K 143460, K 143477, K 147557, K 146913, K 146914, K 147048, TKP2021-NKTA-64, and 2025-1.1.5-NEMZ\_KI-2025-00004, and MATE KKP and KKPCs Research Excellence and Flagship Research Groups grants (Hungary); the Council of Science and Industrial Research, India; ICSC -- National Research Center for High Performance Computing, Big Data and Quantum Computing, FAIR -- Future Artificial Intelligence Research, and CUP I53D23001070006 (Mission 4 Component 1), funded by the NextGenerationEU program, the Italian Ministry of University and Research (MUR) under Bando PRIN 2022 -- CUP I53C24002390006, PRIN PRIMULA 2022RBYK7T (Italy); the Latvian Council of Science; the Ministry of Science and Higher Education, project no. 2022/WK/14, and the National Science Center, contracts Opus 2021/41/B/ST2/01369, 2021/43/B/ST2/01552, 2023/49/B/ST2/03273, and the NAWA contract BPN/PPO/2021/1/00011 (Poland); the Funda\c{c}\~ao para a Ci\^encia e a Tecnologia (Portugal); the National Priorities Research Program by Qatar National Research Fund; MICIU/AEI/10.13039/501100011033, ERDF/EU, ``European Union NextGenerationEU/PRTR", projects PID2022-142604OB-C21, PID2022-139519OB-C21, PID2023-147706NB-I00, PID2023-148896NB-I00, PID2023-146983NB-I00, PID2023-147115NB-I00, PID2023-148418NB-C41, PID2023-148418NB-C42, PID2023-148418NB-C43, PID2023-148418NB-C44, PID2024-158190NB-C22, RYC2021-033305-I, RYC2024-048719-I, CNS2023-144781, CNS2024-154769 and Plan de Ciencia, Tecnolog{\'i}a e Innovaci{\'o}n de Asturias, Spain; the Chulalongkorn Academic into Its 2nd Century Project Advancement Project, the National Science, Research and Innovation Fund program IND\_FF\_68\_369\_2300\_097, and the Program Management Unit for Human Resources \& Institutional Development, Research and Innovation, grant B39G680009 (Thailand); the Eric \& Wendy Schmidt Fund for Strategic Innovation through the CERN Next Generation Triggers project under grant agreement number SIF-2023-004; the Kavli Foundation; the Nvidia Corporation; the SuperMicro Corporation; the Welch Foundation, contract C-1845; and the Weston Havens Foundation (USA).
\end{sloppypar}
\end{acknowledgments}\section*{Data availability} Release and preservation of data used by the CMS Collaboration as the basis for publications is guided by the  \href{https://doi.org/10.7483/OPENDATA.CMS.1BNU.8V1W}{CMS data preservation, re-use and open access policy}.

\bibliography{auto_generated}
\cleardoublepage \appendix\section{The CMS Collaboration \label{app:collab}}\begin{sloppypar}\hyphenpenalty=5000\widowpenalty=500\clubpenalty=5000\cmsinstitute{Yerevan Physics Institute, Yerevan, Armenia}
{\tolerance=6000
A.~Hayrapetyan, V.~Makarenko\cmsorcid{0000-0002-8406-8605}, A.~Tumasyan\cmsAuthorMark{1}\cmsorcid{0009-0000-0684-6742}
\par}
\cmsinstitute{Institut f\"{u}r Hochenergiephysik, Vienna, Austria}
{\tolerance=6000
W.~Adam\cmsorcid{0000-0001-9099-4341}, J.W.~Andrejkovic, L.~Benato\cmsorcid{0000-0001-5135-7489}, T.~Bergauer\cmsorcid{0000-0002-5786-0293}, M.~Dragicevic\cmsorcid{0000-0003-1967-6783}, C.~Giordano\cmsorcid{0000-0001-6317-2481}, P.S.~Hussain\cmsorcid{0000-0002-4825-5278}, M.~Jeitler\cmsAuthorMark{2}\cmsorcid{0000-0002-5141-9560}, N.~Krammer\cmsorcid{0000-0002-0548-0985}, A.~Li\cmsorcid{0000-0002-4547-116X}, D.~Liko\cmsorcid{0000-0002-3380-473X}, M.~Matthewman, I.~Mikulec\cmsorcid{0000-0003-0385-2746}, J.~Schieck\cmsAuthorMark{2}\cmsorcid{0000-0002-1058-8093}, D.~Schwarz\cmsorcid{0000-0002-3821-7331}, R.~Sch\"{o}fbeck\cmsAuthorMark{2}\cmsorcid{0000-0002-2332-8784}, M.~Shooshtari\cmsorcid{0009-0004-8882-4887}, M.~Sonawane\cmsorcid{0000-0003-0510-7010}, W.~Waltenberger\cmsorcid{0000-0002-6215-7228}, C.E.~Wulz\cmsAuthorMark{2}\cmsorcid{0000-0001-9226-5812}
% ...
\par}
$^{1}$Also at Yerevan State University, Yerevan, Armenia\\
$^{2}$Also at Technische Universit\"{a}t Wien, Vienna, Austria\\
% ...
\end{sloppypar}
\end{document}